\pdfoutput=1
\RequirePackage{ifpdf}
\ifpdf % We are running pdfTeX in pdf mode
\documentclass[pdftex]{sigma}
\else
\documentclass{sigma}
\fi

\usepackage[Symbol]{upgreek}

\numberwithin{equation}{section}

\newtheorem{Theorem}{Theorem}[section]
\newtheorem{Corollary}[Theorem]{Corollary}
\newtheorem{Lemma}[Theorem]{Lemma}
\newtheorem{Proposition}[Theorem]{Proposition}
{\theoremstyle{definition}
\newtheorem{Example}[Theorem]{Example}
\newtheorem{Remark}[Theorem]{Remark}
}

\begin{document}

\newcommand{\arXivNumber}{1405.0574}

\allowdisplaybreaks

\renewcommand{\PaperNumber}{096}

\FirstPageHeading

\ShortArticleName{Invariant Poisson Realizations and the Averaging of Dirac Structures}

\ArticleName{Invariant Poisson Realizations\\
and the Averaging of Dirac Structures}

\Author{Jos\'e A.~VALLEJO~$^\dag$ and Yurii VOROBIEV~$^\ddag$}

\AuthorNameForHeading{J.A.~Vallejo and Yu.~Vorobiev}

\Address{$^\dag$~Facultad de Ciencias, Universidad Aut\'onoma de San Luis Potos\'i, M\'exico}
\EmailD{\href{mailto:jvallejo@fc.uaslp.mx}{jvallejo@fc.uaslp.mx}}
\URLaddressD{\url{http://galia.fc.uaslp.mx/~jvallejo/}}

\Address{$^\ddag$~Departamento de Matem\'aticas, Universidad de Sonora, M\'exico}
\EmailD{\href{mailto:yurimv@guaymas.uson.mx}{yurimv@guaymas.uson.mx}}

\ArticleDates{Received May 19, 2014, in f\/inal form September 09, 2014; Published online September 15, 2014}

\Abstract{We describe an averaging procedure on a~Dirac manifold, with respect to a~class of compatible actions of
a~compact Lie group.
Some averaging theorems on the existence of invariant realizations of Poisson structures around (singular) symplectic
leaves are derived.
We show that the construction of coupling Dirac structures (invariant with respect to locally Hamiltonian group actions)
on a~Poisson foliation is related with a~special class of exact gauge transformations.}

\Keywords{Poisson structures; Dirac structures; geometric data; averaging operators}

\Classification{53D17; 70G45; 53C12}

\vspace{-2mm}

\section{Introduction}

Our aim is to discuss some aspects of the averaging procedure on Poisson manifolds which carry singular symplectic
foliations.

Let $(M,\Pi)$ be a~Poisson manifold endowed with a~Poisson tensor~$\Pi$.
The characteristic distribution generated by the Hamiltonian vector f\/ields on $(M,\Pi)$ is integrable in the sense of
Stefan--Sussmann~\cite{Ste-74, Sus-73}, and gives rise to the smooth symplectic foliation $(\mathcal{S},\omega)$, having
a~leaf-wise symplectic form~$\omega$.
The singular situation occurs when there are points where the rank of the Poisson tensor~$\Pi$ is not locally constant.
In this case the leaf-wise symplectic form~$\omega$ has a~singular behavior, in the sense that~$\omega$ can not be
represented as the pull-back of a~global $2$-form on~$M$.
Given a~leaf-preserving (non-canonical) action of a~compact connected Lie group~$G$ on~$M$, we are interested in the
existence of an invariant realization of~$\Pi$ around a~(singular) symplectic leaf~$S$, that is, a~$G$-invariant Poisson
structure $\overline{\Pi}$ which is Poisson-isomorphic to~$\Pi$ on a~neighborhood of~$S$.
Such invariant Poisson realizations appear naturally in the theory of normal forms for Hamiltonian systems of adiabatic
type on general phase spaces~\cite{MJY-13, Vor-11}, which is a~motivation for the present work.

\looseness=-1  Our intention is to describe a~natural reconstruction procedure for a~$G$-invariant Poisson structure $\overline{\Pi}$
from the original one~$\Pi$.
It is clear that, in the contravariant setting, the standard averaging technique~\cite{MaMoRa-90} does not work because
of the nonlinear character of the Jacobi identity.
The alternative proposed here is the construction of an invariant Poisson structure~$\overline{\Pi}$ by applying
averaging arguments to the leaf-wise symplectic form~$\omega$.
In doing so, we must deal with smoothness and non-degeneracy issues, which are not trivial at all in the singular case.
The crucial point is that the smoothness condition for leaf-wise pre-symplectic forms can be formulated within the
category of Dirac structures~\cite{Cou-90, CuWe-86}.
This allows us to develop the reconstruction procedure within the framework of the Dirac formalism, starting with the
Dirac structure $D=\operatorname{Graph}\Pi\subset TM\oplus T^{\ast}M$ associated with~$\Pi$.
We remark that if the $G$-action is compatible in an appropriate way with the leaf-wise pre-symplectic form~$\omega$,
then its $G$-average is a~smooth leaf-wise pre-symplectic form on $\mathcal{S}$ inducing a~$G$-invariant Dirac structure
$\overline{D}$.
Moreover, the Dirac structures~$\overline{D}$ and~$D$ are related by a~gauge transformation~\cite{Bur-05, Bur-Ra-03, SeWe-01} associated to an exact $2$-form on~$M$.
Therefore, by the averaging procedure we mean here the passage from~$D$ to $\overline{D}$.
When a~non-degeneracy condition holds, the Dirac structure~$\overline{D}$ is the graph of a~Poisson bi-vector f\/ield~$\overline{\Pi}$ with the property that $\overline{\Pi}$ is invariant with respect to the $G$-action, and
gauge-equivalent to~$\Pi$.
Combining these arguments with the Moser path method for Poisson structures~\cite{CrMa-13, FMa-13, MZ-06, Vo-05}, we get
that the Poisson structure $\overline{\Pi}$ gives an invariant realization of~$\Pi$ around the symplectic leaf~$S$.

If the manifold~$M$ carries the additional structure of a~regular foliation $\mathcal{F}$, we apply the above results to
the class of $\mathcal{F}$-coupling
Poisson tensors~\cite{Va-04, Vo-01}.
Let $\Pi=\Pi_{2,0}+\Pi_{0,2}$ be a~coupling Poisson tensor on $(M,\mathcal{F})$, where the ``regular part''
$\Pi_{2,0}\in\Gamma(\wedge^{2}\mathbb{H})$ is a~bi-vector f\/ield of constant rank, and the ``singular part''
$\Pi_{0,2}\in \Gamma(\wedge^{2}\mathbb{V})$ is a~leaf-tangent Poisson tensor.
We show that, if the $G$-action is compatible with the singular part $\Pi_{0,2}$ and the symplectic leaf $\mathcal{S}$
of~$\Pi$ is transversal to the foliation $\mathcal{F}$, then~$\Pi$ admits an invariant realization around the leaf which
is again a~$\mathcal{F}$-coupling Poisson structure $\overline{\Pi}=\overline{\Pi}_{2,0}+\overline{\Pi}_{0,2}$, with
$G$-invariant regular and singular parts.
In particular, the compatibility condition automatically holds when the $G$-action is locally Hamiltonian on
$(M,\Pi_{0,2})$.

We also present an alternative approach to the construction of Dirac manifolds with symmetry within the class of
coupling Dirac structures~\cite{BrF,DWa-09,DW,Va-06,Wa}.
Starting with a~Poisson foliation $(M,\mathcal{F},P)$ endowed with the locally Hamiltonian action of a~compact Lie
group~$G$, we describe an averaging procedure $D\mapsto\overline{D}$, for compatible $\mathcal{F}$-coupling Dirac
structures~$D$ on $(M,\mathcal{F},P)$, in terms of the gauge transformations of the corresponding integrable geometric
data~\cite{Vor-08}.
This approach is based on the averaging technique for Poisson connections originally developed, for Hamiltonian group
actions on Poisson f\/iber bundles, in~\cite{MaMoRa-90}.
Here we use a~foliated version of this technique which does not require the existence of a~global momentum map.
With a~dif\/ferent perspective, an averaging procedure was also introduced in~\cite{JoRa-10, JRS-11}, to construct induced
Dirac structures in the context of the reduction method on Dirac manifolds with symmetry.

\looseness=-1
The paper is organized as follows: General properties of averaging operators are reviewed in Section~\ref{sec2}.
In Section~\ref{sec3}, we describe the averaging procedure on Dirac manifolds with respect to a~class of compatible
$G$-actions, and study its relation with exact gauge transformations.
In Section~\ref{sec4}, we formulate and prove the Poisson averaging theorem on the existence of invariant realizations
of Poisson structures around (singular) symplectic leaves.
In Section~\ref{sec5}, within the class of coupling Poisson structures on a~foliated manifold, the $G$-invariant
splittings for Poisson models around a~symplectic leaf transversal to the foliation, are described by using the
bi-graded contravariant calculus and gauge type transformations.
Section~\ref{sec6} is devoted to the study of some symmetries of the structure equations (integrability conditions) for
geometric data on Poisson foliations.
We describe a~class of gauge transformations of integrable geometric data which are equivalent with exact gauge
transformations of Dirac structures preserving the coupling property.
In Section~\ref{sec7}, these results are used for the construction of coupling Dirac structures on a~Poisson foliation,
invariant with respect to locally Hamiltonian $G$-action, in terms of the ``averaged'' integrable geometric data.
We also describe some cohomological obstructions to the construction of Dirac manifolds with Hamiltonian $G$-symmetry in
the context of the averaging procedure.

\vspace{-1mm}

\section{Averaging operators}
\label{sec2}

Let~$G$ be a~compact connected Lie group and $\mathfrak{g}$ its Lie algebra.
Suppose we are given a~smooth (left) action $\Phi:G\times M\rightarrow M$, $(g,m)\mapsto\Phi(g,m)=\Phi_{g}(m)$.
Denote by $a_{M}\in\chi (M)$ the inf\/initesimal generator of~$\Phi$ associated to an element $a\in\mathfrak{g}$,
\begin{gather*}
a_{M}(m)=\left.
\frac{\mathrm{d}}{\mathrm{d}t}\right|_{t=0}\Phi_{\exp(ta)}(m).
\end{gather*}
Let $\mathcal{T}_{s}^{r}(M)$ be the space of all smooth tensor f\/ields on~$M$ of type $(r,s)$.
The $G$-average of every $F\in\mathcal{T}_{s}^{r}(M)$ is a~tensor f\/ield of the same type
$\langle F\rangle^{G}\in\mathcal{T}_{s}^{r}(M)$ given by the formula
\begin{gather}
\langle F\rangle^{G}:=\int_{G}\Phi_{g}^{\ast}F\,\mathrm{d}g,
\label{E1}
\end{gather}
where $\mathrm{d}g$ is the normalized Haar measure on~$G$.
A~tensor f\/ield~$F$ is said to be $G$-invariant if $\Phi_{g}^{\ast}F=F$ for any $g\in G$ or, equivalently,
$\langle F\rangle^{G}=F$.
In inf\/initesimal terms, using the Lie derivative, the $G$-invariance of~$F$ reads ${\mathcal L}_{a_{M}}F=0$ $\forall\,
a\in\mathfrak{g}$.
Moreover, we have the useful identity $\langle{\mathcal L}_{a_{M}}F\rangle^{G}=0$, for every $F\in\mathcal{T}_{s}^{r}(M)$.

Since~$G$ is compact and connected, the exponential mappings $\exp:\mathfrak{g}\rightarrow G$, constructed from the Lie
group structure and from the corresponding bi-invariant Riemannian structure, coincide, and this map is surjective.
Consider the cut locus~$C$ of the identity $e\in G$.
Then,
\begin{gather*}
\exp|_{\mathcal{D}}: \ \mathcal{D}\rightarrow G\setminus C
\end{gather*}
is a~dif\/feomorphism between an open, bounded, star-shaped neighborhood $\mathcal{D}$ of $0\in\mathfrak{g}$, and the
complement $G\setminus C$.
Moreover, $\exp(\partial\mathcal{D})=C$ has zero measure (these are standard results in Riemannian geometry; for
instance, see chapter III in~\cite{Cha-06}).
Let $\mu=\mathrm{d}g$ be the normalized Haar measure on~$G$, considered as a~left (right) volume form on~$G$, and denote
by $\mathrm{Hom}(\mathfrak{g};\mathcal{T}_{s}^{r}(M))$ the space of $\mathbb{R}$-linear mappings
$\lambda:\mathfrak{g\rightarrow}\mathcal{T}_{s}^{r}(M)$.
Then, we can def\/ine an averaging operator $\delta^{G}:\mathrm{Hom}
(\mathfrak{g};\mathcal{T}_{s}^{r}(M))\rightarrow\mathcal{T}_{s}^{r}(M)$ as follows:
\begin{gather}
\label{E2}
\delta^{G}(\lambda):= \int_{\mathcal{D}}\left(\int^1_0\Phi_{\exp(t a)}^{\ast}\lambda_{a}\, \mathrm{d}t
\right)\exp^{\ast}\mu.
\end{gather}

\begin{Example}
Let $G=\mathbb{S}^{1}=\mathbb{R}\setminus2\pi\mathbb{Z}$.
Suppose that an $\mathbb{S}^{1}$-action is generated by the $2\pi$-periodic f\/low of a~vector f\/ield~$\Upsilon$ on~$M$.
Then, formula~\eqref{E2} reads
\begin{gather*}
%\label{E3}
\delta^{G}(\lambda)=-\frac{1}{2\pi}\int_{0}^{2\pi}(t-\pi)(\operatorname{Fl}_{\Upsilon}^{t})^{\ast}F\,\mathrm{d}t+\pi\langle F\rangle^{\mathbb{S}^{1}},
\end{gather*}
where $\lambda=aF$, $F\in\mathcal{T}_{s}^{r}(M)$, and $a\in\mathbb{R}$.
\end{Example}

Consider also the mapping $l^{G}:\mathcal{T}_{s}^{r}(M)\rightarrow \mathrm{Hom}(\mathfrak{g};\mathcal{T}_{s}^{r}(M))$
given~by
\begin{gather*}
l^{G}(F):\mathfrak{g}\ni a\mapsto {\mathcal L}_{a_{M}}F.
\end{gather*}

The following useful fact follows from standard averaging arguments~\cite{MaMoRa-90}.

\begin{Lemma}
The averaging operator $\langle \cdot\rangle^{G}:\mathcal{T}_{s}^{r}(M)\rightarrow \mathcal{T}_{s}^{r}(M)$ has the
representation:
\begin{gather}
\label{AR1}
\langle \cdot\rangle^{G}=\operatorname*{id}+\delta^{G}\circ l^{G}.
\end{gather}
\end{Lemma}
\begin{proof}
Firstly, we have (for any $F\in \mathcal{T}_{s}^{r}(M)$),
\begin{gather}
\label{AR3}
\Phi_{\exp a}^{\ast}F-F=\int_{0}^{1}\Phi_{\exp ta}^{\ast}({\mathcal L}_{a_{M}}F)\,\mathrm{d}t.
\end{gather}
On the other hand, since~$C$ is a~subset of measure~$0$, the average of every $F\in\mathcal{T}_{s}^{r}(M)$ can be
written as
\begin{gather*}
\langle F\rangle^{G}=\int_{\mathcal{D}}(\Phi^\ast_{\exp a}F)\exp^{\ast}\mu.
\end{gather*}
Using this property, and integrating equation~\eqref{AR3} over $\mathcal{D}$, we get~\eqref{AR1}.
\end{proof}

The operators $\langle\cdot\rangle^{G}$, $\delta^{G}$, and $l^{G}$ can be restricted to the spaces of multi-vector f\/ields
$\chi^{k}(M)$ and dif\/ferential forms $\Omega^{k}(M)$.
In particular, it follows from~\eqref{AR1} that the $G$-average of a~closed $k$-form,
$\beta\in\Omega_{\operatorname*{cl}}^{k}(M)$, is
\begin{gather}
\label{AR2}
\langle\beta\rangle^{G}=\beta-\mathrm{d}\circ\delta^{G}(\rho),
\end{gather}
where $\rho\in\mathrm{Hom}(\mathfrak{g};\Omega^{k}(M))$ is given by the insertion operator
$\rho_{a}:=-\mathbf{i}_{a_{M}}\beta$.

\section{Compatible group actions}
\label{sec3}

\subsection{Generalities on Dirac structures}

First, we recall some basic properties of Dirac structures which can be found, for example,
in~\cite{Bur-05,Bur-Ra-03,Cou-90,CuWe-86,JoRa-10}.

A Dirac structure on a~manifold~$M$ is a~smooth distribution $D\subset TM\oplus T^{\ast}M$ which is maximally isotropic
with respect to the natural symmetric pairing
\begin{gather*}
\langle(X,\alpha),(Y,\beta)\rangle:=\beta(X)+\alpha(Y),
\end{gather*}
and involutive with respect to the Courant bracket
\begin{gather*}
\lbrack(X,\alpha),(Y,\beta)]:=\left([X,Y],{\mathcal L}_{X}\beta-{\mathcal L}_{Y}\alpha+\frac{1}{2} \mathrm{d}(\alpha(Y)-\beta(X))\right).
\end{gather*}
Here $(X,\alpha)\in\Gamma(D)$ is a~(local) smooth section of~$D$.
Let $p_{T}:TM\oplus T^{\ast}M\to TM$ be the natural projection.
It follows that $\operatorname*{rank}D=\dim M$, and the characteristic distribution $\mathcal{C}=p_{T}(D)\subset TM$ is
integrable in the sense of Stefan and Sussmann.
As a~consequence, the Dirac manifold $(M,D)$ carries a~(singular) pre-symplectic foliation $(\mathcal{S},\omega)$: the
leaves are maximal integral manifolds of $\mathcal{C}=T\mathcal{S}$, and the leaf-wise pre-symplectic structure~$\omega$
is def\/ined by $\omega_{m}(X,Y)=-\alpha(Y)$, where $(X,Y)\in \mathcal{C}_{m}$ and $(X,\alpha)\in D_{m}$.
In particular, the foliation $(\mathcal{S},\omega)$ is symplectic if and only if~$D$ is the graph of a~Poisson structure
on~$M$.
Reciprocally, one can associate to a~pre-symplectic foliation $(\mathcal{S},\omega)$ on~$M$, the distribution
\begin{gather*}
D^{\omega}:=\{(X,\alpha)\in T_{m}M\oplus T_{m}^{\ast}M\,|\, X\in T_{m}\mathcal{S},
\;
\alpha|_{T_{m}\mathcal{S}}=-\mathbf{i}_{X}\omega_{m}\}.
\end{gather*}
We say that~$\omega$ is smooth if $D^{\omega}$ is a~smooth sub-bundle of $TM\oplus T^{\ast}M$.
In this case, $D^{\omega}$ is a~Dirac structure whose pre-symplectic foliation is just $(\mathcal{S},\omega)$.
Thus, there is a~one-to-one correspondence between Dirac structures and smooth pre-symplectic foliations on~$M$.

Notice that every Poisson structure on~$M$ induces a~Dirac structure for which the involutive property follows from the
Jacobi identity.
Indeed, given a~bi-vector f\/ield $\Pi\in\Gamma(\wedge^{2}TM)$, we def\/ine the smooth sub-bundle
\begin{gather*}
D_{\Pi}:=\operatorname{Graph}(\Pi)=\{(X,\alpha)\in T_{m}M\oplus T_{m}^{\ast}M\,|\, X=\mathbf{i}_{\alpha}\Pi\}.
\end{gather*}
Then, $D_{\Pi}$ is a~Dirac structure if and only if $[\![\Pi,\Pi]\!]=0$, that is, $\Pi$ is a~Poisson bi-vector f\/ield.
Here $[\![\cdot,\cdot]\!]$ denotes the Schouten bracket for multi-vector f\/ields on~$M$~\cite{Va--94}.
In this case, $(\mathcal{S},\omega)$ is in fact the symplectic foliation of the Poisson structure~$\Pi$.
Reciprocally, if the characteristic foliation of a~Dirac structure~$D$ is symplectic, then~$D$ is the graph of a~Poisson
structure.

\looseness=-1
One can modify the leaf-wise pre-symplectic structure~$\omega$ of a~Dirac structure~$D$ by using the pull back of
a~closed 2-form $B\in\Omega^{2}(M)$: for each pre-symplectic leaf $(S,\omega_{S})$, we def\/ine the new pre-symplectic
structure as $\omega_{S}-\iota_{S}^{\ast}B$, where $\iota_{S}:S\hookrightarrow M$ is the inclusion map.
Then, the foliation $\mathcal{S}$ endowed with the deformed leaf-wise pre-symplectic structure gives rise to the new
Dirac structure $\tau_{B}(D)=\{(X,\alpha-\mathbf{i}_{X}B):(X,\alpha)\in D\}$.
Therefore, for every closed 2-form~$B$, the transformation $\tau_{B}$ (called the gauge
transformation~\cite{Bur-Ra-03, SeWe-01}) sends Dirac structures to Dirac structures.

A Dirac structure~$D$ on~$M$ is said to be invariant with respect to a~dif\/feomorphism $\phi:M\rightarrow M$ if
$(\phi^{\ast}X,\phi^{\ast}\alpha)\in\Gamma(D)$ for every $(X,\alpha)\in\Gamma(D)$.
In this case, $\phi$ is called a~Dirac dif\/feomorphism.
In particular, if $D=\operatorname{Graph}\Pi$ is the Dirac structure associated to a~Poisson bi-vector f\/ield~$\Pi$
on~$M$, then the $\phi$-invariance of~$D$ is equivalent to the condition $\phi^{\ast}\Pi=\Pi$, that is, $\phi$ is
a~Poisson dif\/feomorphism.
An action of a~Lie group on $(M,D)$ by Dirac dif\/feomorphisms is called canonical.

A vector f\/ield~$X$ on~$M$ is Hamiltonian relative to a~Dirac structure~$D$ if there exists a~function $F\in
C^{\infty}(M)$ such that
\begin{gather}
\label{H1}
(X,\mathrm{d}F)\in\Gamma(D).
\end{gather}
A~$G$-action on $(M,D)$ by Dirac dif\/feomorphisms is said to be Hamiltonian, with momentum map
$J\in\mathrm{Hom}(\mathfrak{g};C^{\infty}(M))$, if the inf\/initesimal generator $a_{M}$ of every $a\in\mathfrak{g}$ is
a~Hamiltonian vector f\/ield,
\begin{gather*}
%\label{H2}
(a_{M},\mathrm{d}J_{a})\in\Gamma(D).
\end{gather*}
The integrability and reduction for these actions has been studied in~\cite{BrF-13}.

\subsection[Averaging procedure: $D\mapsto\bar{D}$]{Averaging procedure: $\boldsymbol{D\mapsto\bar{D}}$}

Now, let $(M,D)$ be a~Dirac manifold and $(\mathcal{S},\omega)$ the associated pre-symplectic foliation.
Suppose we are given the action of a~connected compact Lie group~$G$ on~$M$ which preserves each leaf of~$\mathcal{S}$
but is not necessarily canonical relative to~$D$.
Therefore, the $G$-action is tangent to the pre-symplectic leaves, $a_{M}(m)\in T_{m}\mathcal{S}$, for all $m\in M$,
$a\in \mathfrak{g}$.
Applying the averaging operator~\eqref{E1} to~$\omega_{S}$ on every pre-symplectic leaf~$(S,\omega_{S})$, gives the
averaged leaf-wise pre-symplectic form~$\langle\omega\rangle^{G}$ on~$\mathcal{S}$.

We say that the \textit{leaf-preserving} $G$-\textit{action} on the Dirac manifold $(M,D)$\ is \textit{compatible}   if
there exists a~$\mathbb{R} $-linear mapping $\rho\in\mathrm{Hom}(\mathfrak{g},\Omega^{1}(M))$ such that, for each leaf~$S$,
\begin{gather}
\label{C1}
\mathbf{i}_{a_{M}}\omega_S =-i^*_S\rho_a,
\end{gather}
for every $a\in\mathfrak{g}$, where $i_S:S\hookrightarrow M$ is the canonical injection.
This compatibility condition can be rewritten as follows
\begin{gather*}
(a_{M},\rho_{a})\in\Gamma(D),
\qquad
\forall\, a\in\mathfrak{g}.
\end{gather*}
It is clear that this condition always holds for Hamiltonian $G$-actions on $(M,D)$, and also when the pre-symplectic
foliation is regular (of constant rank).

\begin{Proposition}
\label{pro3.1}
If the $G$-action is compatible on $(M,D)$, then the average $\langle\omega\rangle^{G}$ is smooth, and can be represented as
\begin{gather}
\label{C2}
\langle\omega\rangle^{G}=\omega-i^*_S\mathrm{d}\Theta,
\end{gather}
where $\Theta\in\Omega^{1}(M)$ is the $1$-form given~by
\begin{gather}
\label{Pr1}
\Theta:=\delta^{G}(\rho).
\end{gather}
The associated Dirac structure $\overline{D}:=D^{\langle\omega\rangle^{G}}$ is $G$-invariant and related to~$D$ by an exact
gauge transformation,
\begin{gather*}
\overline{D}=\{(X,\alpha+\mathbf{i}_{X}\mathrm{d}\Theta):(X,\alpha)\in D\}.
\end{gather*}

\end{Proposition}
\begin{proof}
It follows directly from~\eqref{AR2},~\eqref{C1} and the properties of the gauge transformations.
\end{proof}

\begin{Remark}
The $1$-form~$\Theta$ in~\eqref{C2} is def\/ined up to the addition of an arbitrary $1$-form on~$M$ which is closed on
each leaf~$S$.
It follows from~\eqref{C2}, and the fact that $\langle\cdot\rangle^G$ commutes with pull-backs, that
$\iota_S^{\ast}\langle\mathrm{d}\Theta\rangle^{G}=0$, and hence one can always choose the gauge $1$-form having zero average,
by making $\Theta^{0}=\Theta-\langle\Theta\rangle^{G}$.
\end{Remark}

We will use the following notations.
For an arbitrary bi-vector f\/ield $\Pi\in\Gamma(\wedge^{2}TM)$, and a~$2$-form $B\in \Omega^{2}(M)$, we denote~by
$\Pi^{\sharp}:T^{\ast}M\rightarrow TM$ and $B^{\sharp}:TM\rightarrow T^{\ast}M$ the vector bundle morphisms given~by
$\alpha\mapsto\mathbf{i}_{\alpha}\Pi$ and $X\mapsto\mathbf{i}_{X}B$, respectively.

Now, we formulate the following ``Poisson version'' of Proposition~\ref{pro3.1}.

\begin{Corollary}
Let $(M,\Pi)$ be a~Poisson manifold and $(\mathcal{S},\omega)$ its symplectic foliation.
Suppose that an action of a~compact connected Lie group~$G$ on~$M$ is compatible with the Poisson tensor~$\Pi$, in the
sense that
\begin{gather}
a_{M}=\Pi^{\sharp}\rho_{a},
\qquad
\forall\, a\in\mathfrak{g},
\label{IN1}
\end{gather}
for a~certain $\rho\in\mathrm{Hom}(\mathfrak{g},\Omega^{1}(M))$, and consider the $1$-form~$\Theta$ in~\eqref{Pr1}.
If the endomorphism
\begin{gather}
\label{Pr2}
\operatorname{Id}+(\mathrm{d}\Theta)^{\sharp}\circ\Pi^{\sharp}: \ T^{\ast}M\rightarrow T^{\ast}M \ \ \text{is invertible},
\end{gather}
then the $G$-average $\langle\omega\rangle^{G}$ is non-degenerate on each leaf of $\mathcal{S}$, and there exists a~unique
$G$-inva\-riant Poisson tensor $\overline{\Pi}$ on~$M$ whose singular symplectic foliation is just
$(\mathcal{S},\langle\omega\rangle^{G})$.
The Poisson structures $\overline{\Pi}$ and~$\Pi$ are related by the exact gauge transformation,
\begin{gather}
\label{GT1}
\overline{\Pi}^{\sharp}=\Pi^{\sharp}\circ\big(\operatorname{Id}+(\mathrm{d}\Theta)^{\sharp} \circ\Pi^{\sharp}\big)^{-1}.
\end{gather}
\end{Corollary}

Therefore, under the non-degeneracy condition~\eqref{Pr2}, one can get a~Poisson structure which is invariant with
respect to the compatible $G$-action.

\section{The averaging theorem around symplectic leaves}
\label{sec4}

Here, we apply the results of the previous section to the construction of invariant Poisson models around a~(singular)
symplectic leaf.

\begin{Theorem}
\label{thm4.1}
Let $(M,\Pi)$ be a~Poisson manifold, and~$S$ a~symplectic leaf of the foliation induced by~$\Pi$.
Suppose we are given an action of a~compact connected Lie group~$G$ on~$M$, which is compatible with~$\Pi$
$($recall~\eqref{IN1}$)$.
If~$G$ acts canonically on~$S$ $($but not necessarily on the other leaves$)$, that is, recalling~\eqref{C1}
\begin{gather}
\iota_{S}^{\ast}\rho_{a}
\
\text{is closed on}
\
S,
\label{Dec3}
\end{gather}
then~\eqref{GT1} determines a~$G$-invariant Poisson bi-vector $\overline{\Pi}$, well-defined in a~$G$-invariant
neighborhood~$N$ of~$S$ in~$M$.
Moreover, if
\begin{gather}
S\subset M^{G}
\qquad
(\text{the set of fixed points}),
\label{Dec5}
\end{gather}
then, the germs at~$S$ of $\overline{\Pi}$ and~$\Pi$ are isomorphic by a~local Poisson diffeomorphism
$\phi:N\rightarrow M$,
\begin{gather}
\phi^{\ast}\Pi=\overline{\Pi},
\qquad
\left.\phi\right|_{S}=\mathrm{id}.
\label{PD}
\end{gather}
\end{Theorem}

\begin{proof}
Fix a~sub-bundle, $\mathbb{V}_{S}\subset T_{S}M$, which is transverse to the symplectic leaf,
\begin{gather*}
T_{S}M=TS\oplus\mathbb{V}_{S}.
\end{gather*}
Then,
\begin{gather}
T_{S}^{\ast}M=\mathbb{V}_{S}^{0}\oplus(TS)^{0},
\label{Dec1}
\end{gather}
and it follows from $\Pi^{\sharp}(T_{S}^{\ast}M)=TS$ that
\begin{gather}
\Pi^{\sharp}\big((TS)^{0}\big)=0,
\qquad
\Pi^{\sharp}\big(\mathbb{V}_{S}^{0}\big)=TS.
\label{Dec4}
\end{gather}
Let~$\Theta$ be the 1-from on~$M$ given by~\eqref{Pr1}.
Consider the bundle morphism $B^{\sharp}:TM\rightarrow T^{\ast}M$ induced by the $2$-form $B=-\mathrm{d}\Theta$.
By condition~\eqref{Dec3}, $\Theta$ is closed on~$S$, and hence $\iota_{S}^{\ast}B=0$ or, equivalently,
\begin{gather*}
B^{\sharp}(TS)\subseteq(TS)^{0}.
\end{gather*}
From that result, and properties~\eqref{Dec4}, we get
\begin{gather*}
B^{\sharp}\circ\Pi^{\sharp}\big(\mathbb{V}_{S}^{0}\big)\subseteq(TS)^{0},
\qquad \text{and} \qquad B^{\sharp}\circ\Pi^{\sharp}\big((TS)^{0}\big)=0.
\end{gather*}
These relations, together with~\eqref{Dec1}, mean that, for each $t\in \lbrack0,1]$, the restriction of the vector
bundle morphism $\operatorname{Id} -tB^{\sharp}\circ\Pi^{\sharp}$ to $T_{S}^{\ast}M$ is invertible, with
\begin{gather}
\label{Dec6}
\big(\operatorname{Id}-tB^{\sharp}\circ\Pi^{\sharp}\big)^{-1}=\operatorname{Id}+tB^{\sharp}\circ\Pi^{\sharp}
\qquad
\text{on}
\quad
T_{S}^{\ast}M.
\end{gather}
Then, there exists an open neighborhood~$N$, of~$S$ in~$M$, such that the restriction of
$\operatorname{Id}-tB^{\sharp}\circ\Pi^{\sharp}$ to $T_{N}^{\ast}M$ is invertible for all $t\in [0,1]$.
Since the Lie group~$G$ is compact, one can choose the neighborhood~$N$ as being $G$-invariant.
Applying to~$\Pi$ the gauge transformation determined by~$tB$, we get a~$t$-dependent family of Poisson tensors
$\Pi_{t}$ on~$N$, such that:
\begin{gather*}
\Pi_{t}^{\sharp}:=\Pi^{\sharp}\circ\big(\operatorname{Id}-tB^{\sharp}\circ \Pi^{\sharp}\big)^{-1}.
\end{gather*}
This family joins the original Poison structure, $\Pi$, with the $G$-invariant one $\overline{\Pi}=\Pi_{1}$.
Next, one can verify~\cite{CrMa-13,FMa-13, Vo-01,Vo-05} that the time-dependent vector f\/ield on~$N$ given~by
\begin{gather*}
Z_{t}=-\Pi_{t}^{\sharp}(\Theta)=-\Pi^{\sharp}\circ\big(\operatorname{Id}
+t(\mathrm{d}\Theta)^{\sharp}\circ\Pi^{\sharp}\big)^{-1}(\Theta),
\end{gather*}
satisf\/ies the homotopy equation
\begin{gather*}
[\![Z_{t},\Pi_{t}]\!]
=-\frac{\mathrm{d}\Pi_{t}}{\mathrm{d}t}.
\end{gather*}
Finally, hypothesis~\eqref{Dec5} implies that $\Theta|_{T_{S}M} \in(TS)^{0}$, and hence, by~\eqref{Dec6}, we
get $\left.
Z_{t}\right|_{S}=0$.
Therefore, shrinking if necessary the neighborhood~$N$, we can make the f\/low $\operatorname*{Fl}_{Z_{t}}^{t}$ of $Z_{t}$
well def\/ined on~$N$, for all $t\in [0,1]$.
The Poisson dif\/feomorphism in~\eqref{PD} is then given by the f\/low at time~$1$, $\phi=
\operatorname*{Fl}_{Z_{t}}^{t}|_{t=1}$.
\end{proof}

\section[$G$-invariant splittings]{$\boldsymbol{G}$-invariant splittings}
\label{sec5}

According to the coupling procedure~\cite{Vo-01}, in a~neighborhood of a~closed symplectic leaf, a~Poisson structure
splits into ``regular'' and ``singular'' parts, where the singular part is called a~transverse Poisson structure of the
leaf.
In this section, by using Theorem~\ref{thm4.1}, we show that, with respect to a~class of transversally compatible
$G$-actions, such a~splitting can be made $G$-invariant, and compute the invariant regular and singular components in
terms of gauge transformations.

Let $\mathcal{F}$ be a~\emph{regular} foliation on a~manifold~$M$.
Denote by $\mathbb{V}:=T\mathcal{F}$ the tangent bundle of $\mathcal{F}$, and by $\mathbb{V}^{0}\subset T^{\ast}M$ its
annihilator.
Recall that a~Poisson bi-vector f\/ield, $\Pi\in\Gamma(\wedge^{2}TM)$, on the foliated manifold $(M,\mathcal{F})$ is said
to be $\mathcal{F}$-\emph{coupling}~\cite{Va-04, Vo-01}, if the associated distribution
\begin{gather*}
\mathbb{H}:=\Pi^{\sharp}\big(\mathbb{V}^{0}\big),
\end{gather*}
is a~normal (regular) bundle of $\mathcal{F}$, that is,
\begin{gather*}
TM=\mathbb{H}\oplus\mathbb{V},
%\label{Spl1}
\end{gather*}
and hence
\begin{gather*}
T^{\ast}M=\mathbb{V}^{0}\oplus\mathbb{H}^{0}.
%\label{Spl2}
\end{gather*}
These splittings def\/ine an $\mathbb{H}$-dependent bi-grading of dif\/ferential forms and multi-vector f\/ields on~$M$:
\begin{gather*}
\Omega^{k}(M)=  \bigoplus_{p+q=k}\Omega^{p,q}(M),
\qquad
\Gamma\big({\wedge}^{k}TM\big)=  \bigoplus_{p+q=k}\chi^{p,q}(M),
\end{gather*}
where the elements of the sub-spaces $\Omega^{p,q}(M)=\Gamma(\wedge^{q}\mathbb{V}^{0}\otimes\wedge^{p}\mathbb{H}^{0})$,
and $\chi^{p,q}(M)=\Gamma(\wedge^{p}\mathbb{H}\otimes\wedge^{q}\mathbb{V})$, are said to be dif\/ferential forms and
multi-vector f\/ields of bi-degree $(p,q)$, respectively.
For any $k$-form~$\omega$, and $k$-vector f\/ield~$A$, the terms of bi-degree $(p,q)$ in the above decompositions will be
denoted by $\omega_{p,q}$ and $A_{p,q}$, respectively.
Moreover, we will use the following bi-graded decomposition of the exterior dif\/ferential, $\mathrm{d}$,
on~$M$~\cite{Va--94,Va-04}:
\begin{gather}
\label{DEC}
\mathrm{d}=\mathrm{d}_{1,0}+\mathrm{d}_{2,-1}+\mathrm{d}_{0,1}.
\end{gather}

For every $\mathcal{F}$\textit{-}coupling Poison tensor~$\Pi$, the mixed term $\Pi_{1,1}$, of bi-degree $(1,1)$,
vanishes and we have the decomposition
\begin{gather*}
\Pi=\Pi_{2,0}+\Pi_{0,2},
\end{gather*}
where the ``regular part'', $\Pi_{2,0}\in\Gamma(\wedge^{2}\mathbb{H})$ is a~bi-vector f\/ield of constant rank,
\begin{gather}
\operatorname{rank}\Pi_{2,0}=\dim\mathbb{H=}\operatorname{codim}\mathcal{F},
\label{D1}
\end{gather}
and the ``singular part'', $\Pi_{0,2}\in\Gamma(\wedge^{2}\mathbb{V})$, is a~leaf-wise tangent Poisson tensor,
\begin{gather*}
\Pi_{0,2}^{\sharp}(T^{\ast}M)\subset\mathbb{V},
\qquad
\text{and}
\qquad
[\![\Pi_{0,2},\Pi_{0,2}]\!]=0.
\end{gather*}
It follows from~\eqref{D1} that the restriction of $\Pi_{2,0}^{\sharp}$ to $\mathbb{V}^{0}$ is a~vector bundle
isomorphism onto $\mathbb{H}$,
\begin{gather*}
\Pi_{2,0}^{\sharp}\big(\mathbb{V}^{0}\big)=\mathbb{H}.
\end{gather*}
For every $1$-form $\beta=\beta_{1,0}+\beta_{0,1}$, where $\beta_{1,0}\in \Gamma(\mathbb{V}^{0})$ and
$\beta_{0,1}\in\Gamma(\mathbb{H}^{0})$, we have
\begin{gather}
\label{ID1}
\Pi^{\sharp}\beta=\Pi_{2,0}^{\sharp}\beta_{1,0}+\Pi_{0,2}^{\sharp}\beta_{0,1}.
\end{gather}
Therefore, the characteristic distribution of~$\Pi$ is the direct sum of the normal bundle $\mathbb{H}$, and the
characteristic distribution of $\Pi_{0,2}$,
\begin{gather*}
\Pi^{\sharp}(T^{\ast}M)=\mathbb{H}\oplus\Pi_{0,2}^{\sharp}\big(\mathbb{H}^{0}\big).
\end{gather*}
This shows that the sets of singular points of the Poisson structures~$\Pi$ and $\Pi_{0,2}$ coincide.
Moreover, the symplectic leaves of~$\Pi$ intersect the leaves of $\mathcal{F}$ transversally and symplectically.
Notice also that $\Pi_{2,0}$ is a~Poisson tensor if and only if the distribution $\mathbb{H}$ is integrable.

Now, given an $\mathcal{F}$-coupling Poisson structure $\Pi=\Pi_{2,0}+\Pi_{0,2}$ on $(M,\mathcal{F})$, we will assume
that the action of a~compact connected Lie group~$G$ on~$M$ is def\/ined, such that it is compatible with the leaf-wise
tangent Poisson tensor $\Pi_{0,2}$ in the sense that
\begin{gather}
a_{M}=\Pi_{0,2}^{\sharp} \mu_{a} ,
\qquad
\forall\,
a\in\mathfrak{g},
\label{GA1}
\end{gather}
for a~certain $\mu=\mu_{1,0}+\mu_{0,1}\in\mathrm{Hom}(\mathfrak{g},\Omega^{1}(M))$.
Let
\begin{gather*}
\Theta:=\delta^{G}(\mu_{0,1}),
%\label{GA2}
\end{gather*}
and consider the $2$-forms on~$M$
\begin{gather*}
%\label{GA3}
B:=-\mathrm{d}\Theta,
\qquad
B_{0,2}=-\mathrm{d}_{0,1}\Theta_{0,1}.
\end{gather*}

\begin{Theorem}
\label{thm5.1}
Let $S\subset M$ be a~symplectic leaf of~$\Pi$, such that
\begin{gather}
\label{TR1}
T_{S}M=TS\oplus T_{S}\mathcal{F}.
\end{gather}
Then, in a~$G$-invariant open neighborhood, $N$, of~$S$ in~$M$, the Poisson tensor~$\Pi$ is isomorphic to
a~$\mathcal{F}$-coupling Poisson tensor $\overline{\Pi}=\overline{\Pi}_{2,0}+\overline{\Pi}_{0,2}$, whose regular and
singular components
\begin{gather}
\overline{\Pi}_{2,0}
\
\mbox{and}
\
\overline{\Pi}_{0,2}
\
\text{are $G$-invariant}.
\label{TR2}
\end{gather}
Around the leaf~$S$, the Poisson structures $\overline{\Pi}$ and $\overline{\Pi}_{0,2}$ are related with~$\Pi$ and
$\Pi_{0,2}$ by the gauge transformations
\begin{gather}
 \overline{\Pi}^{\sharp}=\Pi^{\sharp}\circ\big(\operatorname{Id}-B^{\sharp}\circ \Pi^{\sharp}\big)^{-1},
\label{TR3}
\\
 \overline{\Pi}_{0,2}^{\sharp}=\Pi_{0,2}^{\sharp}\circ\big(\operatorname{Id} -B_{0,2}^{\sharp}\circ\Pi_{0,2}^{\sharp}\big)^{-1}.
\label{TR4}
\end{gather}
\end{Theorem}

\begin{proof}
It follows from~\eqref{ID1} and~\eqref{GA1} that
\begin{gather*}
a_{M}=\Pi_{0,2}^{\sharp}(\mu_{a})_{0,1}=\Pi^{\sharp}(\mu_{a})_{0,1}
\end{gather*}
and hence condition~\eqref{IN1} holds for~$\rho=\mu_{0,1}$.
Therefore, the $G$-action is also compatible with~$\Pi$.
The transversality condition~\eqref{TR1} says that $\Pi_{0,2}$ vanishes on~$S$, and hence $\left.
a_{M}\right|_{S}=0$.
Then, by Theorem~\ref{thm4.1}, in a~$G$-invariant neighborhood~$N$ of~$S$, the gauge transformation~\eqref{TR3}
determines the $G$-invariant Poisson tensor $\overline{\Pi}$, which is isomorphic to~$\Pi$ by a~local
dif\/feomorphism~$\phi$ which restricts to the identity on~$S$.
Since the characteristic distributions of~$\Pi$ and $\overline{\Pi}$ coincide on~$N$, we conclude that~$S$ is
a~symplectic leaf of $\overline{\Pi}$.
Again by~\eqref{TR1}, one can choose the neighborhood~$N$ (shrinking it, if needed) in such a~way that
$\overline{\mathbb{H}}:=\overline{\Pi}^{\sharp}(\mathbb{V}^{0})$ is a~normal bundle of $\mathcal{F}$, and hence
$\overline{\Pi}=\overline{\Pi}_{2,0}+\overline{\Pi}_{0,2}$ is a~$\mathcal{F}$-coupling Poisson tensor on~$N$ (see, for
example~\cite{Va-04}).
It follows from~\eqref{GA1} that $a_{M}\in\Gamma(\mathbb{V})$, and hence we have the inclusions
\begin{gather*}
 [\![\overline{\Pi}_{2,0},a_{M}]\!]\in\chi^{2,0}(M)\oplus\chi^{1,1}(M),
\qquad
 [\![\overline{\Pi}_{0,2},a_{M}]\!]\in\chi^{0,2}(M),
\end{gather*}
where the bi-grading is taken with respect to the decomposition $TM=\mathbb{\bar{H}}\oplus\mathbb{V}$.
These pro\-per\-ties, and the $G$-invariance of $\overline{\Pi}$, imply that $[\![\overline{\Pi}_{2,0},a_{M}]\!]=[\![\overline{\Pi}_{0,2},a_{M}]\!]=0$, for every $a\in\mathfrak{g}$.
This proves~\eqref{TR2}.
Now, let us check~\eqref{TR4}.
Consider the projection $\bar{p}_{V}:TM\rightarrow\mathbb{V}$, along $\mathbb{\bar{H}}$.
Equality~\eqref{TR3} reads
\begin{gather*}
\overline{\Pi}_{2,0}^{\sharp}+\overline{\Pi}_{0,2}^{\sharp}-\overline{\Pi}_{2,0}^{\sharp}\circ
B^{\sharp}\circ\Pi^{\sharp}-\overline{\Pi}_{0,2}^{\sharp}\circ
B^{\sharp}\circ\Pi^{\sharp}=\Pi_{2,0}^{\sharp}+\Pi_{0,2}^{\sharp},
\end{gather*}
but taking into account the properties $\overline{\Pi}_{2,0}^{\sharp}(T^{\ast}M)=\mathbb{\bar{H}}$, and
$\overline{\Pi}_{0,2}^{\sharp}(T^{\ast}M)\subset\mathbb{V}$, this equality is equivalent to the following relations,
involving $\bar{p}_{V}$,
\begin{gather*}
 \overline{\Pi}_{2,0}^{\sharp}-\overline{\Pi}_{2,0}^{\sharp}\circ B^{\sharp}\circ
\Pi^{\sharp}=(\operatorname*{id}-\bar{p}_{V})\circ\Pi_{2,0}^{\sharp},
\qquad
 \overline{\Pi}_{0,2}^{\sharp}-\overline{\Pi}_{0,2}^{\sharp}\circ B^{\sharp}\circ
\Pi^{\sharp}=\bar{p}_{V}\circ\Pi_{2,0}^{\sharp}+\Pi_{02}^{\sharp}.
\end{gather*}
As $\Pi_{2,0}^{\sharp}(\mathbb{H}^{0})=0$, and
$\overline{\Pi}_{0,2}^{\sharp}(\mathbb{V}^{0})=\Pi_{0,2}^{\sharp}(\mathbb{V}^{0})=0$, we conclude that the last equality
splits into the following:
\begin{gather*}
  \overline{\Pi}_{0,2}^{\sharp}-\overline{\Pi}_{0,2}^{\sharp}\circ B^{\sharp}\circ
\Pi_{0,2}^{\sharp}=\Pi_{0,2}^{\sharp},
\qquad
  \overline{\Pi}_{0,2}^{\sharp}\circ B^{\sharp}\circ\Pi_{2,0}^{\sharp}=-\bar{p}_{V}\circ\Pi_{2,0}^{\sharp}.
\end{gather*}
By decomposing $B=B_{2,0}+B_{1,1}+B_{0,2}$, and using the properties $B_{2,0}^{\sharp}(\mathbb{V})=0$, and
$B_{1,1}^{\sharp} (\mathbb{V})\subset\mathbb{V}^{0}$, we get $\overline{\Pi}_{0,2}^{\sharp}
\circ(B_{2,0}+B_{1,1})^{\sharp}\circ\Pi_{0,2}^{\sharp}=0$.
Hence $\bar{\Pi}_{0,2}^{\sharp}-\overline{\Pi}_{0,2}^{\sharp}\circ
B_{0,2}^{\sharp}\circ\Pi_{0,2}^{\sharp}=\Pi_{02}^{\sharp}$.
To f\/inish, it suf\/f\/ices to notice that $\operatorname{Id}+B_{0,2}^{\sharp}\circ\Pi_{0,2}^{\sharp}$ is invertible around
the leaf~$S$, because of the property $\Pi_{0,2}=0$ at~$S$.
\end{proof}

\begin{Corollary}
The regular component of $\overline{\Pi}$ is given~by
\begin{gather}
\overline{\Pi}_{2,0}^{\sharp}=(\operatorname*{id}-\bar{p}_{V})\circ\Pi_{2,0}^{\sharp}\circ\big(\operatorname{Id}-B^{\sharp}\circ\Pi^{\sharp}\big)^{-1}.
\label{AL}
\end{gather}
\end{Corollary}

\begin{Remark}
If the distribution $\mathbb{H}$ is integrable, then $\overline{\Pi}$ splits into two $G$-invariant Poisson structures:
$\overline{\Pi}_{2,0}$, and $\overline{\Pi}_{0,2}$.
Locally, around the f\/ixed points of canonical group actions, such splitting always exists due to the equivariant
versions of Weinstein splitting theorem~\cite{FMa-13, MZ-06}.
\end{Remark}

Now, consider the case of a~$G$-action which is locally Hamiltonian on $(M,\Pi_{0,2})$, that is, the compatibility
condition~\eqref{GA1} holds for a~certain $\mu\in\mathrm{Hom}(\mathfrak{g},\Omega^{1}_{\mathrm{cl}}(M))$.
That means $\mathrm{d}\mu_{a}=0$, for every $a\in\mathfrak{g}$, and the inf\/initesimal generator $a_{M}$ is locally
Hamiltonian vector f\/ield on $(M,\Pi_{0,2})$.
Then, $0=(\mathrm{d}\mu_{a})_{0,2}=\mathrm{d}_{0,1}(\mu_{a})_{0,1}$, and hence $B_{0,2}=0$.
Notice that the operator $\delta^{G}$ is compatible with the f\/iltration given by $\Omega^{p,\bullet}$, so
\begin{gather*}
\delta^{G}\big(\mathrm{Hom}\big(\mathfrak{g},\Omega^{p,q}(M)\big)\big) \subset \bigoplus_{k\geq 0}\Omega^{p+k,q-k}(M).
\end{gather*}
Moreover, since $\delta^{G}$ commutes with the exterior derivative, the $1$-form $\delta^{G}(\mu_{a})$ is closed, and
the gauge $2$-form in~\eqref{TR3} can be represented as $B:=-\mathrm{d}Q$, where
\begin{gather*}
Q:=-\delta^{G}(\mu_{1,0})\in\Omega^{1,0}(M)=\Gamma\big(\mathbb{V}^{0}\big).
\end{gather*}
Here we are using the property $\delta^{G}(\mathrm{Hom}(\mathfrak{g},\Omega^{p,0}(M)))\subset\Omega^{p,0}(M)$.
It follows that $B=B_{2,0}+B_{1,1}$, with
\begin{gather*}
B_{2,0}=\mathrm{d}_{1,0}\circ\delta^{G}(\mu_{1,0}),
\qquad \text{and} \qquad
B_{1,1}=\mathrm{d}_{0,1}\circ\delta^{G}(\mu_{1,0}).
\end{gather*}
Thus, in this case Theorem~\ref{thm5.1} guarantees that, around the symplectic leaf~$S$, $\Pi$ is Poisson-dif\/feomorphic
to the $G$-invariant $\mathcal{F}$-coupling Poisson tensor $\overline{\Pi}$, with $\overline{\Pi}_{0,2}=\Pi_{0,2}$.

In particular, if the action of the Lie group~$G$ on $(M,\Pi_{0,2})$ is Hamiltonian with momentum map
$J\in\mathrm{Hom}(\mathfrak{g},C^{\infty}(M))$, so
\begin{gather*}
a_{M}=\Pi_{0,2}^{\sharp}\mathrm{d}J_{a},
\qquad
\forall\,
a\in\mathfrak{g},
\end{gather*}
then
\begin{gather*}
%\label{HC1}
Q=-\delta^{G}(\mathrm{d}_{1,0}J).
\end{gather*}
For example, in the case $G=\mathbb{S}^{1}=\mathbb{R}\setminus 2\pi\mathbb{Z}$, we have:
\begin{gather*}
Q=\frac{1}{2\pi}\int_{0}^{2\pi}(t-\pi)\big(\mathrm{Fl}_{\Pi_{0,2}^{\sharp}\mathrm{d}J}^{t}\big)^{\ast}\mathrm{d}_{1,0}J\,
\mathrm{d}t-\pi\langle\mathrm{d}_{1,0}J\rangle.
\end{gather*}

The adiabatic situation described in the following example, typically occurs in the theory of perturbations of
Hamiltonian systems~\cite{MJY-13, Vor-11}.

\begin{Example}\looseness=-1
Let~$M$ be a~connected symplectic manifold (viewed as a~parameter space), and let $\mathcal{P}$ be a~Poisson manifold
endowed with a~smooth family of locally Hamiltonian actions $\Phi^m:\mathcal{P}\times G\rightarrow\mathcal{P}$ (where
$m\in M$), of a~compact connected Lie group~$G$.
Let $x_{0}\in \mathcal{P}^{G}$ be a~f\/ixed point at which the Poisson structure on $\mathcal{P}$ has zero rank.
Then, around the slice $M\times\{x_{0}\}$ (considered as a~singular symplectic leaf), the product Poisson structure on
$M\times\mathcal{P}$ is Poisson equivalent to the $G$-invariant Poisson tensor which gives rise to the averaged
Hamiltonian dynamics.
\end{Example}

\section{Gauge transformations of geometric data}
\label{sec6}

In this section we describe a~class of exact gauge transformations of coupling Dirac structures on a~foliated manifold
which preserve the coupling property.

\subsection{Connections on foliated manifolds}

Suppose we have a~regular foliated manifold $(M,\mathcal{F})$.
Let be $\mathbb{V}=T\mathcal{F}$ the tangent bundle, also called the vertical distribution.
Recall that a~vector valued $1$-form $\gamma\in\Omega^{1}(M;\mathbb{V})$ is a~connection on $(M,\mathcal{F})$ if the
vector bundle morphism $\gamma:TM\rightarrow\mathbb{V}$ satisf\/ies the projection property $\gamma\circ\gamma=\gamma$,
and $\operatorname{Im}\gamma=\mathbb{V}$.
Then, $\mathbb{H}:=\ker\gamma$ is a~normal bundle of $\mathcal{F}$, called the horizontal sub-bundle (with respect to
the leaf space $M\setminus\mathcal{F}$).
Reciprocally, given a~normal bundle $\mathbb{H}$ of $\mathcal{F}$, one can def\/ine the associated connection as the
projection $\gamma=p_{V}:TM\rightarrow\mathbb{V}$, according to the decomposition $TM=\mathbb{H}\oplus\mathbb{V}$.

The curvature of a~connection~$\gamma$ is the vector valued $2$-form $R^{\gamma}\in\Omega^{2}(M;\mathbb{V})$ on~$M$
given by $R^{\gamma}=\frac{1}{2} [\gamma,\gamma]_{\operatorname*{FN}}$.
Here $[,]_{\mathrm{FN}}:\Omega^{k}(M;TM)\times\Omega^{l}(M;TM)\rightarrow\Omega^{k+l}(M;TM)$ denotes the
Fr\"{o}licher--Nijenhuis bracket~\cite{KMS-93} of vector-valued forms on~$M$.
For example, for any $K,L\in\Omega^{1}(M;TM)$, we have
\begin{gather}
\lbrack K,L]_{\operatorname*{FN}}(X,Y) =[KX,LY]-[KY,LX]-L\left([KX,Y]-[KY,X]\right)
\nonumber
\\
\phantom{\lbrack K,L]_{\operatorname*{FN}}(X,Y)=}{}
-K\left([LX,Y]-[LY,X]\right) +(LK+KL)[X,Y],
\label{FN}
\end{gather}
where $X,Y\in\mathfrak{X}(M)$.

Recall also that a~vector f\/ield~$X$ on~$M$ is said to be projectable (on the leaf space $M\setminus\mathcal{F}$) if
$[X,\Gamma(\mathbb{V})]\subset\Gamma (\mathbb{V})$.
The space of all (local) projectable vector f\/ields is denoted by $\chi_{\operatorname*{pr}}(M,\mathcal{F})$.
For a~given connection~$\gamma$, by $\Gamma_{\operatorname*{pr}}(\mathbb{H})$ we denote the set of all (local)
projectable sections of the horizontal subbundle $\mathbb{H}$.
Then, the spaces $\chi(M)$ and $\Gamma(\mathbb{H})$ are locally generated by the elements of
$\chi_{\operatorname*{pr}}(M,\mathcal{F})$ and $\Gamma_{\operatorname*{pr}}(\mathbb{H})$, respectively.
In particular, the curvature of a~connection~$\gamma$ is uniquely determined by the relations
\begin{gather*}
R^{\gamma}(X,Y)=\gamma([X,Y]),
\qquad
\forall\,
X,Y\in\Gamma_{\operatorname*{pr}}(\mathbb{H}),
\end{gather*}
and $\mathbf{i}_{V}R^{\gamma}=0$ for all $V\in\Gamma(\mathbb{V})$.

Fix a~connection~$\gamma$; then, any other connection $\tilde{\gamma}$ is of the form $\tilde{\gamma}=\gamma-\Xi$, where
$\Xi\in\Omega^{1} (M;\mathbb{V})$ is a~vector valued 1-form satisfying the condition
$\mathbb{V\subseteq}\operatorname{Ker}\Xi$.
The horizontal subbundle of $\tilde{\gamma}$ can be represented as
\begin{gather}
\mathbb{\tilde{H}}=\ker\tilde{\gamma}=(\operatorname{Id}+\Xi)(\mathbb{H}).
\label{HS}
\end{gather}
Moreover, the transition rule for the curvature form reads
\begin{gather}
R^{\tilde{\gamma}}=R^{\gamma}-\left([\gamma,\Xi]_{\operatorname*{FN}} -\frac{1}{2}[\Xi,\Xi]_{\operatorname*{FN}}\right).
\label{Cur}
\end{gather}

Suppose now that the foliated manifold $(M,\mathcal{F})$ is endowed with a~leaf-wise tangent Poisson bi-vector f\/ield
$P\in\Gamma(\wedge^{2}\mathbb{V})$.
Then, each leaf of $\mathcal{F}$ inherits a~Poisson structure from~$P$ and we have a~Poisson foliation denoted~by
$(M,\mathcal{F},P)$.
A~\textit{connection}~$\gamma$ is said to be \textit{Poisson} on $(M,\mathcal{F},P)$ if every projectable section
$X\in\Gamma_{\operatorname*{pr}}(\mathbb{H})$ of the horizontal bundle $\mathbb{H}$ is a~Poisson vector f\/ield on~$(M,P)$.
In this case, for every $X\in\Gamma_{\operatorname*{pr}}(\mathbb{H})$, $R^{\gamma}(X,Y)$ is a~vertical Poisson vector
f\/ield.

\subsection{Coupling Dirac structures}

By a~set of \emph{geometric data} on a~foliated manifold $(M,\mathcal{F})$, we mean a~triple $(\gamma,\sigma,P)$
consisting of a~connection $\gamma\in\Omega^{1}(M;\mathbb{V})$, a~horizontal $2$-form
$\sigma\in\Gamma(\wedge^{2}\mathbb{V}^{0})$ on~$M$, and a~leaf-wise tangent Poisson tensor
$P\in\Gamma(\wedge^{2}\mathbb{V})$.
The geometric data $(\gamma,\sigma,P)$ are said to be integrable if they satisfy the \emph{structure equations}
\begin{gather}
  {\mathcal L}_{X}P=0,
\label{SE1}
\\
  \mathrm{d}_{1,0}^{\gamma}\sigma=0,
\label{SE2}
\\
  R^{\gamma}(X,Y)=-P^{\sharp}\mathrm{d}\sigma(X,Y),
\label{SE3}
\end{gather}
for any $X,Y\in\Gamma_{\operatorname*{pr}}(\mathbb{H})$.
Here $\mathbb{H} =\ker\gamma$ is the horizontal sub-bundle and $\mathrm{d}_{1,0}^{\gamma}$ is the operator of bi-degree
$(1,0)$, in the decomposition~\eqref{DEC}, associated to $\mathbb{H}$.
In particular, one has
\begin{gather}
\label{6.6prima}
\mathrm{d}_{1,0}^{\gamma}\beta(X_{0},X_{1},\dots ,X_{q})= \mathrm{d}\beta(X_{0},X_{1},\dots ,X_{q})
\end{gather}
for any $\beta\in\Gamma(\wedge^{q}\mathbb{V}^{0}\mathcal{)}$, and
$X_{0},X_{1},\dots ,X_{q}\in\Gamma_{\operatorname*{pr}}(\mathbb{H})$.
Here $\mathrm{d}$ is the exterior dif\/ferential on~$M$.
Conditions~\eqref{SE1} and~\eqref{SE2} say that~$\gamma$ is a~Poisson connection on $(M,\mathcal{F},P)$, whose curvature
takes values in the vertical Hamiltonian vector f\/ields.

As is known~\cite{Va-04, Vo-01}, every $\mathcal{F}$-coupling Poisson structure~$\Pi$ on $(M,\mathcal{F})$ is
equivalent to a~set of integrable geometric data $(\gamma,\sigma,P)$, such that the restriction of~$\sigma$ to
$\mathbb{H}$ is non-degenerate, that~is,
\begin{gather}
\sigma^{\sharp}|_{\mathbb{H}}:\ \mathbb{H}\rightarrow\mathbb{V}^{0}
\
\text{is invertible}.
\label{ND}
\end{gather}
The bi-vector f\/ield~$\Pi$ can be reconstructed from $(\gamma,\sigma,P)$ by means of the formula
$\Pi=\Pi_{2,0}+\Pi_{0,2}$, where $\Pi_{0,2}=P$, and $\Pi_{2,0}\in \Gamma(\wedge^{2}\mathbb{H})$ is uniquely determined
by the relation $\Pi_{2,0}^{^{\sharp}}|_{\mathbb{V}^{0}}=-(\sigma^{\sharp}|_{\mathbb{H}})^{-1}$.
Therefore, the structure equations~\eqref{SE1},~\eqref{SE2},~\eqref{SE3}, give a~factorization of the Jacobi identity
for~$\Pi$.

A Dirac structure $D\subset TM\oplus T^{\ast}M$ is said to be $\mathcal{F}$-coupling~\cite{Va-06} if the associated
tangent distribution $\mathbb{H}=\mathbb{H}(D,\mathcal{F})$,
\begin{gather}
\mathbb{H}_{m}:=\big\{Z\in T_{m}M:\exists\,\alpha\in\mathbb{V}^{0}
\;
\text{and}
\;
(Z,\alpha)\in D\big\},
\label{TD}
\end{gather}
is a~normal bundle of $\mathcal{F}$.
By lifting the non-degeneracy condition~\eqref{ND}, we get the following fact~\cite{DW,Va-06,Wa}: There exists
a~one-to-one correspondence $(\gamma,\sigma,P)\mapsto D$, between integrable geometric data and $\mathcal{F}$-coupling
Dirac structures on $(M,\mathcal{F})$, which is given~by
\begin{gather*}
D=\big\{\big(X+P^{\sharp}\alpha,\alpha-\mathbf{i}_{X}\sigma\big): X\in\Gamma (\mathbb{H}),\alpha\in\Gamma\big(\mathbb{H}^{0}\big) \big\}
\end{gather*}
or, equivalently,
\begin{gather*}
D=\mathrm{Graph}(
\sigma|_{\mathbb{H}}) \oplus \mathrm{Graph}\big(P|_{\mathbb{H}^0}\big).
\end{gather*}
The leaf-wise pre-symplectic structure associated to an $\mathcal{F}$-coupling Dirac structure~$D$, can be described in
terms of the corresponding geometric data as follows: Recall that the characteristic distribution $p_{T}(D)$, of~$D$, is
integrable, and gives rise to the singular pre-symplectic foliation $(\mathcal{S},\omega)$, where~$\omega$ is
a~leaf-wise pre-symplectic form.
Then, $\mathcal{F}\cap\mathcal{S}$ is a~symplectic foliation of~$P$, and we have
\begin{gather}
T\mathcal{S}=\mathbb{H}\oplus P^{\sharp}\big(\mathbb{V}^{0}\big).
\label{PSF1}
\end{gather}
This implies the point-wise splitting
\begin{gather}
\omega_{m}=\sigma_{m}\oplus\tau_{m},
\qquad
\forall\,
m\in M,
\label{PSF2}
\end{gather}
where~$\tau$ is the leaf-wise symplectic form associated to~$P$.
It follows that $T\mathcal{S}\cap\mathbb{V}$ is the characteristic distribution of~$P$, and in terms of the
pre-symplectic form, the characteristic sub-bundle of the $\mathcal{F}$-coupling Dirac structure~$D$, is represented as
\begin{gather*}
\mathbb{H}_{m}= (T_{m}\mathcal{S}\cap\mathbb{V}_{m} )^{\omega}\equiv\big\{X\in
T_{m}\mathcal{S}:\, \omega_{m}(X,P^{\sharp}df)=0,\; \forall\, f\in C_{\operatorname*{loc}}^{\infty}(M)\big\}.
\end{gather*}

It is useful to rewrite condition~\eqref{H1} (for a~vector f\/ield~$X$ on~$M$ to be Hamiltonian, relative to the
$\mathcal{F}$-coupling Dirac structure~$D$) in terms of the geometric data $(\gamma,\sigma,P)$.
It easy to see that the vector f\/ield $X=X_{1,0}+X_{0,1}$ is Hamiltonian on $(M,D)$ if and only if the components
$X_{1,0}\in\Gamma(\mathbb{H})$, and $X_{0,1}\in\Gamma(\mathbb{V})$, satisfy the relations:
\begin{gather}
X_{0,1}=   P^{\sharp}\mathrm{d}F,
\label{FH1}
\\
\mathbf{i}_{X_{1,0}}\sigma=  -\mathrm{d}_{1,0}^{\gamma}F,
\label{FH2}
\end{gather}
for a~certain $F\in C^{\infty}(M)$.

We remark that there is a~natural class of coupling Dirac structures on vector bundles, which comes from transitive Lie
algebroids and plays an important r\^ole in constructing linearized models around (pre)\,symplectic leaves of Poisson
and Dirac manifolds~\cite{CrMa-13,Va-06, Vo-01}.

\subsection[$Q$-gauge transformations]{$\boldsymbol{Q}$-gauge transformations}

Here, we will describe some symmetries of the structure equations (see also~\cite{Vo-05}).
Let $(\gamma,\sigma,P)$ be some geometric data on $(M,\mathcal{F})$ and $Q\in\Gamma(\mathbb{V}^{0}\mathcal{)}$
a~horizontal $1$-form.
For every $\beta\in\Gamma(\wedge^{q}\mathbb{V}^{0}\mathcal{)}$, denote by $\{Q\wedge\beta\}_{P}$ the element of
$\Gamma(\wedge^{q+1}\mathbb{V}^{0}\mathcal{)}$ given~by
\begin{gather*}
\{Q\wedge\beta\}_{P}(X_{0},X_{1},\dots ,X_{q}):=\sum\limits_{i=0}^{q}(-1)^{i}\big\{Q(X_{i}),\beta\big(X_{0},X_{1},\dots,\hat{X}_{i},\dots,X_{q}\big)\big\}_{P},
\end{gather*}
where $\{f_{1},f_{2}\}_{P}=P(\mathrm{d}f_{1},\mathrm{d}f_{2})$ is the Poisson bracket associated to~$P$.
Def\/ine
\begin{gather}
\tilde{\gamma}:=   \gamma-\Xi^{Q},
\label{T1}
\\
\tilde{\sigma}:=  \sigma-\left(\mathrm{d}_{1,0}^{\gamma}Q+\frac{1}{2}\{Q\wedge Q\}_{P}\right),
\label{T2}
\end{gather}
where $\Xi^{Q}\!\in\!\Omega^{1}(M;\mathbb{V})$ is the vector-valued $1$-form uniquely determined by the condition
$\Xi^{Q}(X)=P^{\sharp}\mathrm{d}Q(X)$, for every $X\in\chi_{\operatorname*{pr}}(M,\mathcal{F})$.
Evidently, the vector-valued $1$-form $\tilde{\gamma}$ determines a~connection on $(M,\mathcal{F})$, and
$\tilde{\sigma}\in\Gamma(\wedge^{2}\mathbb{V}^{0})$.
One can think of the mapping $(\gamma,\sigma,P)\mapsto(\tilde{\gamma},\tilde {\sigma},P)$ as a~gauge transformation
def\/ined on the set of all geometric data on $(M,\mathcal{F})$, leaving f\/ixed the Poisson tensor~$P$.
The following result shows that such gauge transformations preserve the coupling property.

\begin{Proposition}
Let~$D$ be a~$\mathcal{F}$-coupling Dirac structure, associated to the integrable geometric data $(\gamma,\sigma,P)$ on
$(M,\mathcal{F})$, and let $Q\in\Gamma(\mathbb{V}^{0})$ be an arbitrary horizontal $1$-form on~$M$.
Then, the triple $(\tilde{\gamma},\tilde{\sigma},P)$ defined by~\eqref{T1},~\eqref{T2}, satisfies the structure
equations~\eqref{SE1} to~\eqref{SE3}.
Moreover, the $\mathcal{F}$-coupling Dirac structure $\tilde{D}$, associated to the integrable geometric data
$(\tilde{\gamma},\tilde{\sigma},P)$, is related to~$D$ by the exact gauge transformation:
\begin{gather}
\tilde{D}=\big\{(X,\alpha-\mathbf{i}_{X}\mathrm{d}Q): (X,\alpha)\in D\big\}.
\label{Gau}
\end{gather}
\end{Proposition}

\begin{proof}
Let $\mathbb{\tilde{H}=}\ker\tilde{\gamma}$ be the horizontal bundle of $\tilde{\gamma}$.
From~\eqref{HS} and~\eqref{T1}, we get that every projectable vector f\/ield
$\tilde{X}\in\Gamma_{\operatorname*{pr}}(\mathbb{\tilde{H}})$ can be represented as
\begin{gather}
\tilde{X}=X+P^{\sharp}\mathrm{d}Q(X),
\qquad
X\in\Gamma_{\operatorname*{pr}}(\mathbb{H}),
\label{PV}
\end{gather}
and hence $\tilde{X}$ is a~Poisson vector f\/ield with respect to~$P$.
That the curvature identity~\eqref{SE3} for $R^{\tilde{\gamma}}$ is satisf\/ied, can be straightforwardly checked,~by
using the fact that~$\gamma$ is a~Poisson connection, the equality
$\gamma(\tilde{X})=\Xi^{Q}(\tilde{X})=P^{\sharp}\mathrm{d}Q(X)$, and relations~\eqref{FN},~\eqref{Cur}.
The corresponding coupling form $\tilde{\sigma}$ is just given by~\eqref{T2}.
The structure equations for $(\gamma,\sigma,P)$ imply the following identities:
\begin{gather*}
  (\mathrm{d}_{1,0}^{\gamma})^{2}Q=\{Q\wedge\sigma\}_{P},
%\label{TF1}
\qquad
  \mathrm{d}_{1,0}^{\gamma}\{Q\wedge Q\}_{P}= -2\{Q\wedge \mathrm{d}_{1,0}^{\tilde{\gamma}}Q\}_{P}.
%\label{TF2}
\end{gather*}
Moreover, by~\eqref{T1}, we have
\begin{gather*}
\mathrm{d}_{1,0}^{\tilde{\gamma}}\beta= \mathrm{d}_{1,0}^{\gamma}\beta+\{Q\wedge\beta\}_{P},
\qquad
\beta\in\Gamma\big(\wedge^{q}\mathbb{V}^{0}\big).
\end{gather*}
Using these relations, it can be readily checked that $\mathrm{d}_{1,0}^{\tilde{\gamma}}\tilde{\sigma}=0$.
This proves the integrability of $(\tilde{\gamma},\tilde{\sigma},P)$.
Now, consider the Dirac structure $\tilde{D}$, induced by $(\tilde{\gamma},\tilde{\sigma},P)$.
Relations~\eqref{PSF1} and~\eqref{PV}, show that $p_{T}(\tilde{D})=p_{T}(D)$.
Let $(\mathcal{S},\tilde{\omega})$ and $(\mathcal{S},\omega)$ be the pre-symplectic foliations associated to $\tilde{D}$
and~$D$, respectively.
Then, $T\mathcal{S}$ is generated by local projectable vector f\/ields of the form~\eqref{PV}, and
$P^{\sharp}\mathrm{d}f$, where $X\in\Gamma_{\operatorname*{pr}}(\mathbb{H})$, and $f\in
C_{\operatorname*{loc}}^{\infty}(M)$.
Evaluating the pre-symplectic forms $\tilde{\omega}$ and~$\omega$ on this family of vector f\/ields, and using the
point-wise splitting~\eqref{PSF2} for $\tilde{\omega}$, we can verify, by a~straightforward computation, that
\begin{gather}
\label{PRS}
\tilde{\omega}_{S}+\iota_{S}^{\ast}\mathrm{d}Q=\omega_{S}
\end{gather}
at every pre-symplectic leaf~$S$ of $\mathcal{S}$.
This means that $\tilde{D}$ is given by~\eqref{Gau}.
\end{proof}

Therefore, gauge transformations of integrable geometric data lead to exact gauge transformations of Dirac structures.
The reciprocal is also true.

\begin{Proposition}
\label{pro6.2}
For every $Q\in\Gamma(\mathbb{V}^{0}\mathcal)$ and an $\mathcal{F}$-coupling Dirac structure~$D$, the exact gauge
transformation~\eqref{Gau} takes~$D$ to the $\mathcal{F}$-coupling Dirac structure $\tilde{D}$, whose geometric data are
given by~\eqref{T1},~\eqref{T2}.
\end{Proposition}

\begin{proof}
Let us show f\/irst that $\tilde{D}$ is $\mathcal{F}$-coupling.
The Dirac structures $\tilde{D}$ and~$D$ determine the same leaf partition $\mathcal{S}$ of~$M$, and the corresponding
pre-symplectic structures $\tilde{\omega}$ and~$\omega$, are related by~\eqref{PRS}.
Because of~\eqref{PSF1}, any vector f\/ield $X\in\Gamma(\mathbb{H})$ and Hamiltonian vector f\/ield $P^{\sharp}\mathrm{d}f$,
are tangent to the foliation $\mathcal{S}$, and $\omega$-orthogonal.
Then, any arbitrary projectable vector f\/ield $\tilde{X}$, of the form~\eqref{PV}, and $P^{\sharp}\mathrm{d}f$ are
$\tilde{\omega}$-orthogonal,
\begin{gather*}
\tilde{\omega}\big(\tilde{X},P^{\sharp}\mathrm{d}f\big)  = \tau\big(P^{\sharp}\mathrm{d}Q(X),P^{\sharp}\mathrm{d}f\big)-
\mathrm{d}Q\big(\tilde{X},P^{\sharp}\mathrm{d}f\big)
  =\{Q(X),f\}_{P}+{\mathcal L}_{P^{\sharp}\mathrm{d}f}Q(X)=0.
\end{gather*}
According to~\eqref{PV}, the tangent distribution~\eqref{TD}, associated to $\tilde{D}$, is given~by
\begin{gather}
\mathbb{\tilde{H}=}\mathrm{Span}\big\{\tilde{X}=X+P^{\sharp}\mathrm{d}Q(X): X\in\Gamma_{\operatorname*{pr}}(\mathbb{H})\big\},
\label{SAN}
\end{gather}
and hence it is a~normal bundle of $\mathcal{F}$.
Therefore, $\tilde{D}$ is a~$\mathcal{F}$-coupling Dirac structure.
Let $(\tilde{\gamma},\tilde{\sigma},\tilde{P})$ be the corresponding integrable geometric data.
The connection $\tilde{\gamma}$, induced by $\mathbb{\tilde{H}}$, is given by~\eqref{T1}.
Moreover, by~\eqref{PRS} and the condition that~$Q$ is horizontal, we conclude that the restriction of $\tilde{\omega}$
to $T_{m}\mathcal{S}\cap\mathbb{V}_{m}$ coincides with $\tau_{m}$.
Thus, $\tilde{P}=P$.
Finally, using~\eqref{PRS} and~\eqref{SAN}, we compute the coupling $2$-form $\tilde{\sigma}$
\begin{gather*}
\tilde{\sigma}\big(\tilde{X}_{1},\tilde{X}_{2}\big)
= \tilde{\omega}\big(\tilde{X}_{1},\tilde{X}_{2}\big)
=\omega(\tilde{X}_{1},\tilde{X}_{2})-\mathrm{d}Q\big(\tilde{X}_{1},\tilde{X}_{2}\big)
\\
\phantom{\tilde{\sigma}\big(\tilde{X}_{1},\tilde{X}_{2}\big)}
{}  =\sigma(X_{1},X_{2})+\{Q(X_{1}),Q(X_{2})\}_{P}
  -\mathrm{d}Q(X_{1},X_{2})-2\{Q(X_{1}),Q(X_{2})\}_{P}
\\
\phantom{\tilde{\sigma}\big(\tilde{X}_{1},\tilde{X}_{2}\big)}
 {} =\sigma(X_{1},X_{2})-\mathrm{d}Q(X_{1},X_{2})-\{Q(X_{1}),Q(X_{2})\}_{P}.
\end{gather*}
Therefore, $\tilde{\sigma}$ is just given by~\eqref{T2}.
\end{proof}

\begin{Remark}
Gauge transformations of the form~\eqref{T1},~\eqref{T2}, appear naturally in the classif\/ication theory of Poisson
structures around a~symplectic leaf~\cite{Vo-05}, and in the gauge theory on principal bundles~\cite{DWa-09}.
\end{Remark}

\section{Averaging of coupling Dirac structures}
\label{sec7}

\looseness=-1
In this section we present a~generalization of some results obtained in~\cite{Vor-08} in the case of Hamiltonian actions
on Poisson f\/iber bundles.
This time, without the requirement of the existence of a~global momentum map, we describe the averaging procedure for
coupling Dirac (not just Poisson) structures on a~foliated manifold with respect to a~class of locally Hamiltonian group
actions.

Suppose we have an action $\Phi:G\times M\rightarrow M$, of a~compact connected Lie group~$G$ on a~foliated manifold
$(M,\mathcal{F})$, which preserves the foliation, $(\Phi_{g})_{*m}\mathbb{V}_{m}=\mathbb{V}_{\Phi_{g}(m)}$, for all
$g\in G$.
It is clear the pull-back $\Phi_{g}^{\ast}$ preserves the subspaces $\Omega^{p}(M;\mathbb{V})\subset\Omega^{p}(M;TM)$,
and hence the averaging operator $\langle\cdot\rangle^{G}:\Omega^{p}(M;\mathbb{V})\rightarrow \Omega^{p}(M;\mathbb{V})$ is
well-def\/ined on vector valued forms through
\begin{gather*}
\langle K\rangle^{G}(X)=\int_{G}\Phi_{g}^{\ast}(K((\Phi_{g})_{\ast}X)\mathrm{d}g,
\end{gather*}
for every $K\in\Omega^{p}(M;\mathbb{V})$, $X\in\chi(M)$.
Notice that~$K$ is $G$-invariant if and only if $K=\langle K\rangle^{G}$.

In particular, by averaging a~connection~$\gamma$ we obtain a~$G$-invariant connection $\langle\gamma\rangle^{G}$.
Indeed, taking into account that the $G$-action preserves the subspace of vertical vector f\/ields, it easy to see that
$\langle\gamma\rangle^{G}(V)=V$, for all $V\in\Gamma(\mathbb{V})$.
The dif\/ference vector $1$-form $\Xi:=\gamma-\langle\gamma\rangle^{G}\in\Omega^{1}(M;\mathbb{V})$ has zero average,
$\langle\Xi\rangle^{G}=0$, and admits the representation $\Xi=\delta^{G}\circ l^{G}(\gamma)$.
Here the $\mathbb{R}$-linear mapping $l^{G}:\Omega^{1}(M;\mathbb{V})\rightarrow\mathrm{Hom}(\mathfrak{g};\Omega^{1}
(M;\mathbb{V}))$ is def\/ined by $l^{G}(\gamma)_{a}=[a_{M},\gamma]_{\mathrm{FN}}$.
The horizontal bundle $\mathbb{\bar{H}}=(\operatorname{Id}+\Xi)(\mathbb{H})$ of $\langle\gamma\rangle^{G}$, and the curvature
form $R^{\langle\gamma\rangle^{G}}$, are also $G$-invariant.

\subsection[$G$-invariant integrable geometric data]{$\boldsymbol{G}$-invariant integrable geometric data}

As we have seen in Sections~\ref{sec2} and~\ref{sec6}, the averaging procedure for Dirac structures is well-def\/ined with
respect to the class of compatible compact group actions, and is related to the existence exact gauge transformations.
Here we show that the $G$-average $\overline{D}=D^{\langle\omega\rangle^{G}}$ of a~$\mathcal{F}$-coupling Dirac structure~$D$,
with respect to a~locally Hamiltonian $G$-action, inherits the coupling property and give computational formulae for the
corresponding invariant geometric data.

First, we observe that given a~foliation-preserving action $\Phi:G\times M\rightarrow M$, of a~Lie group~$G$ on
$(M,\mathcal{F})$, we have an induced $G$-action on the set of all geometric data on $(M,\mathcal{F})$, def\/ined by the
transformations
\begin{gather*}
(\gamma,\sigma,P)\mapsto\big(\Phi_{g}^{\ast}\gamma,\Phi_{g}^{\ast} \sigma,\Phi_{g}^{\ast}P\big).
\end{gather*}
It is easy to see that these transformations are symmetries of the structures equations~\eqref{SE1} to~\eqref{SE3}.
In other words, the induced action preserves the subset of integrable geometric data.
Recall that a~Dirac structure~$D$ is $G$-invariant if, for any $(\alpha,X)\in\Gamma(D)$ and $g\in G$, we have
$(\Phi_{g}^{\ast}\alpha,\Phi_{g}^{\ast}X)\in\Gamma(D)$.
Then, it is possible to show that an $\mathcal{F}$-coupling Dirac structure is invariant, with respect to the $G$-action
on $(M,\mathcal{F})$, if and only if the associated integrable geometric data $(\gamma,\sigma,P)$ are $G$-invariant,
that is, invariant with respect to the induced $G$-action~\cite{Va-06, Vo-05}.
\begin{Theorem}
\label{thm7.1}
Let~$D$ be a~$\mathcal{F}$-coupling Dirac structure on $(M,\mathcal{F})$, associated to the integrable geometric data
$(\gamma,\sigma,P)$.
Let $\Phi:G\times M\rightarrow M$ be a~locally Hamiltonian action of a~compact connected Lie group~$G$ on
$(M,\mathcal{F},P)$,
\begin{gather}
a_{M}=P^{\sharp}\mu_{a},
\qquad
\mathrm{d}\mu_{a}=0.
\label{Com}
\end{gather}
Then, the $G$-average $\overline{D}=D^{\langle\omega\rangle^{G}}$ of~$D$, is an $\mathcal{F}$-coupling Dirac structure on
$(M,\mathcal{F})$, associated to the $G$-invariant geometric data $(\overline{\gamma},\overline{\sigma},P)$,
\begin{gather}
\overline{D}=\mathrm{Graph}\big(\overline{\sigma}|_{\mathbb{\overline{H}}}\big)\oplus\mathrm{Graph}\big(P|_{\mathbb{\overline{H}}^{0}}\big),
\label{DIR}
\end{gather}
which are given~by
\begin{gather}
\overline{\gamma}:=\langle\gamma\rangle^{G}\equiv\gamma-\Xi^{Q},
\label{OB3}
\\
\overline{\sigma}:=\langle\sigma\rangle^{G}+\frac{1}{2}\langle \{Q\wedge Q\}_{P}\rangle^{G}-
\mathrm{d}_{1,0}^{\overline{\gamma}}\langle Q\rangle^{G}.
\label{OB1}
\end{gather}
Here
\begin{gather}
Q=-\delta^{G}(\mu_{1,0})\in\Gamma\big(\mathbb{V}^{0}\big).
\label{OB2}
\end{gather}
\end{Theorem}

\begin{proof}
It follows from~\eqref{Com} that the locally Hamiltonian $G$-action is compatible with~$D$, and hence,~by
Proposition~\ref{pro3.1}, the average $D^{\langle\omega\rangle^{G}}$ is well-def\/ined and related to~$D$ by the exact gauge
transformation~\eqref{Gau}, where the horizontal $1$-form~$Q$ is given by~\eqref{OB2}.
Then, by Proposition~\ref{pro6.2}, $D^{\langle\omega\rangle^{G}}$ is an $\mathcal{F}$-coupling Dirac structure associated to
the geometric data $(\overline{\gamma},\overline{\sigma},P)$, where $\overline{\gamma}$ is given by~\eqref{OB3}, and
\begin{gather*}
\overline{\sigma}:= \sigma-\left(\mathrm{d}_{1,0}^{\gamma}Q+\frac{1}{2}\{Q\wedge Q\}_{P}\right).
%\label{OB4}
\end{gather*}
Since the averaged Dirac structure is invariant with respect to the $G$-action, the data
$(\overline{\gamma},\overline{\sigma})$ are also $G$-invariant.
Averaging~\eqref{OB3}, and the identity
\begin{gather*}
\big(\mathrm{d}_{1,0}^{\overline{\gamma}}Q\big)= \mathrm{d}_{1,0}^{\gamma}Q+\{Q\wedge Q\}_{P},
\end{gather*}
we get the relations
\begin{gather*}
\overline{\sigma}=\langle\overline{\sigma}\rangle^G= \langle\sigma\rangle^G-\big\langle\mathrm{d}_{1,0}^{\gamma}Q\big\rangle^G-
\frac{1}{2}\langle\{Q\wedge Q\}_{P}\rangle^G,
\end{gather*}
and
\begin{gather*}
\big\langle\mathrm{d}_{1,0}^{\gamma}Q\big\rangle^{G}= \mathrm{d}_{1,0}^{\overline{\gamma}}\langle Q\rangle^{G}+\langle\{Q\wedge Q\}_{P}\rangle^{G}.
\end{gather*}
This proves~\eqref{OB1}.
\end{proof}

As a~consequence of this result, we have the following alternative version of Theorem~\ref{thm5.1}.

\begin{Corollary}
Under the hypotheses of Theorem~{\rm \ref{thm7.1}}, suppose that $D=\mathrm{Graph}\Pi$ is the graph of a~Poisson tensor~$\Pi$ on~$M$,
which has a~symplectic leaf~$S$ satisfying the transversality condition~\eqref{TR1}.
Then, in a~neighborhood of~$S$, we have $\overline{D}=\mathrm{Graph}\overline{\Pi}$, where
$\overline{\Pi}=\overline{\Pi}_{2,0}+P$ is a~$G$-invariant coupling Poisson tensor, whose geometric data are given
by~\eqref{OB3} and~\eqref{OB1}.
In particular, the $G$-invariant component $\overline{\Pi}_{2,0}$ is defined by $($cf.~\eqref{AL}$)$:
\begin{gather*}
\overline{\Pi}_{2,0}^{^{\sharp}}|_{\mathbb{V}^{0}}=-\big(\bar{\sigma}^{\sharp}|_{\overline{\mathbb{H}}}\big)^{-1}.
\end{gather*}
\end{Corollary}

In terms of the geometric data, a~Poisson dif\/feomorphism~$\phi$, between the Poisson structures $\overline{\Pi}$
and~$\Pi$, can be constructed in the following way~\cite{Vo-05}: Consider the family of integrable geometric data
$(\gamma_{t},\sigma_{t},P)$, def\/ined by $\gamma_{t}=\gamma-t\Xi^{Q}$, and
\begin{gather*}
\sigma_{t}= \sigma-\left(t\mathrm{d}_{1,0}^{\gamma}Q+\frac{t^{2}}{2}\{Q\wedge Q\}_{P} \right).
\end{gather*}
Because of the transversality condition, in a~neighborhood of~$S$, $\sigma_{t}|_{\mathbb{H}_{t}}$ is non-degenerate
for all $t\in [0,1]$.
As a~consequence, there exists a~unique time-dependent vector f\/ield $Z_{t}\in \Gamma(\mathbb{H}_{t})$ satisfying the
equation $\mathbf{i}_{Z_{t}}\sigma_{t}=Q$.
Then, $\phi$ is def\/ined by evaluating the f\/low of $Z_{t}$ at time $t=1$.

\subsection[Invariant sections of $\overline{D}$]{Invariant sections of $\boldsymbol{\overline{D}}$}

This brief subsection is devoted to some remarks about invariant sections of the averaged Dirac structure~$\overline{D}$.
First, notice that the $G$-invariant sections of the horizontal bundle $\overline{\mathbb{H}}$ of the averaged Poisson
connection $\overline{\gamma}$~\eqref{OB3}, can be described in the following way.
Let $X\in\Upgamma_{\operatorname*{pr}}(\mathbb{H})$ be a~projectable section of $\mathbb{H}$, def\/ined on an invariant domain of~$M$.
Then, the $G$-average $\langle X \rangle^G$ is a~projectable section of $\overline{\mathbb{H}}$, of the form
\begin{gather}
\label{eq7.7prima}
\langle X\rangle ^G =X+P^\sharp \mathrm{d}Q(X)\in \Upgamma_{\operatorname*{pr}}(\overline{\mathbb{H}}),
\end{gather}
where~$Q$ is given by~\eqref{OB2}.
It follows that
\begin{gather}
\label{invsec1}
\big(X+P^\sharp \mathrm{d}Q(X),-\mathbf{i}_X\overline{\sigma}\big)
\end{gather}
is a~$G$-invariant section of the Dirac structure $\overline{D}$.
Moreover, the sub-bundle $\overline{\mathbb{H}}^0\subset T^*M$ is invariant under the action of~$G$, and every
$G$-invariant $1$-form $\beta\in\Upgamma (\overline{\mathbb{H}}^0)$ induces the $G$-invariant section
\begin{gather}
\label{invsec2}
(P^\sharp \beta,\beta)
\end{gather}
of $\overline{D}$.
However, notice that in general these sections do not generate $\overline{D}$; in the following subsection we will
consider a~case where they do.
On the other hand, it can be shown~\cite{Sni-13} that the Dirac structure $\overline{D}$ is locally spanned~by
$G$-invariant sections; this fact is based on the tube theorem and the averaging procedure for proper Lie group
actions~\cite{JoRa-10,JRS-11}.

\subsection{Hamiltonian actions}

Below we will assume that the hypotheses of Theorem~\ref{thm7.1} hold, and the foliation $\mathcal{F}$ is a~f\/ibration.
Therefore, the leaf space $B=M\setminus\mathcal{F}$ is a~smooth manifold and the natural projection $\pi:M\rightarrow B$
is a~submersion, which we will assume has connected f\/ibers.
In this case, every projectable section $X\in\Gamma_{\operatorname*{pr}}(\mathbb{H})$ is the $\gamma$-horizontal lift of
a~smooth vector f\/ield on~$B$ and, hence, it is well-def\/ined on a~$G$-invariant open domain of~$M$.
This implies the following important property: The horizontal bundle $\overline{\mathbb{H}}$ of $\overline{\gamma}$ is
spanned by $G$-invariant Poisson vector f\/ields of the form~\eqref{OB2}.
As a~consequence, we also get that the averaged Dirac structure $\overline{D}$ is spanned by $G$-invariant sections of
the form~\eqref{invsec1} and~\eqref{invsec2}.

Now, suppose that the action of the Lie group~$G$ on $(M,P)$ is Hamiltonian, with momentum map
$J\in\mathrm{Hom}(\mathfrak{g};C^{\infty}(M))$,
\begin{gather*}
a_{M}=P^{\sharp}\mathrm{d}J_{a},
\qquad
\forall\, a\in\mathfrak{g}.
\end{gather*}
In general, the $G$-action is not Hamiltonian with respect to the original coupling Dirac structure~$D$.
As it follows from~\eqref{FH1},~\eqref{FH2}, this happens only in the particular case $\mathrm{d}_{1,0}^{\gamma}J=0$.
Thus, it is natural to ask whether the $G$-action is Hamiltonian with respect to the averaged Dirac structure
$\overline{D}$~\eqref{DIR}.
The key property in this regard is that, for every $X\in\Gamma_{\operatorname*{pr}}(\mathbb{H})$, we have
\begin{gather}
\langle {\mathcal L}_{X}J_{a}\rangle^{G}={\mathcal L}_{\langle X\rangle^{G}}J_{a}\in\mathrm{Casim}(M;P),
\label{Vaz}
\end{gather}
where $\mathrm{Casim}(M;P)$ denotes the space of Casimir functions of~$P$.
Indeed, noticing f\/irst that
\begin{gather*}
P^{\sharp}\mathrm{d}\big({\mathcal L}_{\langle X\rangle^{G}}J_{a}\big)= \big[\langle X\rangle^{G},P^{\sharp}\mathrm{d}J_{a}\big] =0,
\end{gather*}
we can use~\eqref{eq7.7prima} to get
\begin{gather*}
{\mathcal L}_{\langle X\rangle^{G}}J_{a}={\mathcal L}_{X}J_{a}+\{Q(X),J_{a}\}_{P}={\mathcal L}_{X}J_{a}-{\mathcal L}_{a_{M}}Q(X).
\end{gather*}
Averaging this equality, and taking into account that every Casimir function is $G$-invariant, we obtain
\begin{gather*}
{\mathcal L}_{\langle X \rangle^{G}}J_{a}   = \big\langle {\mathcal L}_{\langle X \rangle^{G}}J_{a}\big\rangle^{G}=
\langle {\mathcal L}_{X}J_{a}\rangle^{G}-\langle {\mathcal L}_{a_{M}}Q(X)\rangle^{G}
  =\langle {\mathcal L}_{X}J_{a}\rangle^{G}.
\end{gather*}
This proves~\eqref{Vaz}.
Moreover, since the group~$G$ acts along the f\/ibers of the projection~$\pi$, the averaging operator preserves the
subspace of horizontal $1$-forms.
In particular, $\langle\mathrm{d}_{1,0}^{\gamma}J_{a}\rangle^{G}\in\Gamma(\mathbb{V}^{0})$.
Using this fact and properties~\eqref{6.6prima},~\eqref{eq7.7prima} and~\eqref{Vaz}, we get
\begin{gather}
\mathbf{i}_{\langle X \rangle^{G}}\mathrm{d}_{1,0}^{\overline{\gamma}}J_{a}
=\mathbf{i}_{\langle X \rangle^{G}}\mathrm{d}^{\overline{\gamma}}J_{a} =\mathbf{i}_{\langle X \rangle^{G}}\mathrm{d}J_{a}
  ={\mathcal L}_{\langle X \rangle^{G}}J_{a} =\langle{\mathcal L}_{X}J_{a}\rangle^{G}
\nonumber
\\
\phantom{\mathbf{i}_{\langle X \rangle^{G}}\mathrm{d}_{1,0}^{\overline{\gamma}}J_{a}}
  =\big\langle\mathbf{i}_{\langle X \rangle^{G}}\mathrm{d}_{1,0}^{\gamma}J_{a}\big\rangle^{G}
=\mathbf{i}_{X}\big\langle\mathrm{d}_{1,0}^{\gamma}J_{a}\big\rangle^{G},\label{Vaz1}
\end{gather}
for every $X\in\Gamma_{\operatorname*{pr}}(\mathbb{H})$.

In this way, we arrive at the following criterion.

\begin{Proposition}
The $G$-action
is Hamiltonian on the Dirac manifold $(M,\overline{D})$, with momentum map~$J$,
\begin{gather*}
(a_{M},\mathrm{d}J_{a})\in\Gamma(\overline{D}),
\qquad
\forall\, a\in\mathfrak{g},
%\label{AD1}
\end{gather*}
if and only if
\begin{gather}
\big\langle \mathrm{d}_{1,0}^{\gamma}J\big\rangle^{G}=0.
\label{AD2}
\end{gather}
\end{Proposition}

\begin{proof}
It follows from~\eqref{FH1},~\eqref{FH2}, that the inf\/initesimal generator $a_{M}$ of the $G$-action is Hamiltonian,
relative to the Dirac structure $\overline{D}$ and the function $J_{a}$, if and only if
$\mathrm{d}_{1,0}^{\overline{\gamma}}J_{a}=0$.
This equality, together with~\eqref{Vaz1}, implies~\eqref{AD2}.
\end{proof}

Therefore, condition~\eqref{AD2} (known as the ``adiabatic hypothesis'' in the theory of Hannay--Berry connections on
Poisson f\/iber bundles~\cite{MaMoRa-90}), appears in the context of the Hamiltonization of the $G$-action with respect to
the Dirac structure $\overline{D}$.
The freedom in the choice of the momentum map is given by the transformation $J\mapsto J-K$, for arbitary
$K\in\mathrm{Hom}(\mathfrak{g};\mathrm{Casim}(M;P))$.
Fixing~$J$, the point is to choose~$K$ in such a~way that $J-K$ satisf\/ies~\eqref{AD2}.
This question can be reformulated in cohomological terms.
Let $\mathcal{C}_{B}^{k}:=\Omega^{k}(B)\otimes\mathrm{Casim}(M;P)$ be the space of $k$-forms on the base~$B$, with
values in the algebra of Casimir functions of the Poisson structure~$P$.
Consider the operator $\partial:\mathcal{C}_{B}^{k}\rightarrow\mathcal{C}_{B}^{k+1}$ def\/ined~by
$(\partial\beta)(u_{1},\dots ,u_{k})= \mathrm{d}(\pi^{\ast}\beta)(X_{1},\dots ,X_{k})$.
Here, $X_{i}\in\Gamma_{\operatorname*{pr}}(\mathbb{H})$ is the $\gamma$-horizontal lift of a~vector f\/ield $u_{i}$ on~$B$, and
$\pi^{\ast}:\mathcal{C}_{B}^{k}\rightarrow\Gamma(\wedge^{k} \mathbb{V}^{0})$ is the pull-back.
Then, it follows from the curvature identity~\eqref{SE3} that $\partial$ is a~coboundary operator,
$\partial^{2}=0$~\cite{Vo-05}.
Property~\eqref{Vaz} implies
\begin{gather*}
\big\langle\mathrm{d}_{1,0}^{\gamma}J\big\rangle^{G}=\pi^{\ast}\zeta
\end{gather*}
for a~certain $\zeta\in\mathrm{Hom}(\mathfrak{g};\mathcal{C}_{B}^{1})$.
Moreover one can prove~\cite{MaMoRa-90} that, for all $X,Y\in\Gamma_{\operatorname*{pr}}(\mathbb{H})$,
\begin{gather*}
\mathbf{i}_{X}\mathbf{i}_{Y}\mathrm{d}\big\langle\mathrm{d}_{1,0}^{\gamma}J_{a}\big\rangle^{G}=0,
\end{gather*}
and hence $\zeta_{a}\in\mathcal{C}_{B}^{k}$ is a~$\partial$-closed $1$-form, for every $a\in\mathfrak{g}$.
Finally, we conclude that the $G$-action admits a~momentum map satisfying~\eqref{AD2} if and only if the
$\partial$-cohomology class of~$\zeta$ is trivial.

\begin{Example}
Consider the particular case in which the Poisson bundle $(\pi:M\to B,P)$, associated to the original coupling Dirac
structure~$D$, is a~locally trivial Lie--Poisson bundle over~$B$ with typical f\/iber $\mathfrak{h}^*$, the dual of a~Lie
algebra $\mathfrak{h}$.
Therefore, the restriction of~$P$ to each f\/iber of~$\pi$ is isomorphic to the Lie--Poisson structure on the co-algebra
$\mathfrak{h}^*$.
Assume also that the Poisson connection~$\gamma$ is homogeneous in the sense that the Lie derivative along every
$X\in\Upgamma_{\operatorname*{pr}}(\mathbb{H})$, preserves the space of f\/iber-wise linear functions $C^\infty_{\rm lin}(M)$ on the total
space of~$\pi$.
Finally, suppose that the momentum map~$J$ of the Hamiltonian $G$-action on $(M,P)$, is f\/iber-wise linear, $J\in
C^\infty_{\rm lin}(M)$.
It follows that, for every $a\in\mathfrak{h}$, the $1$-form $\zeta_a$ on~$B$ takes values in the f\/iber-wise linear
Casimir functions of~$P$.
Our remark is that if the center is trivial, $Z(\mathfrak{h})=\{0\}$, then $\zeta =0$, and hence the $G$-action is
Hamiltonian relative to the Dirac structure $\overline{D}$, with the same momentum map~$J$.

Notice also that, in the case in which the Casimir functions of~$P$ can be described as pull-backs of functions on the
base, $\mathrm{Casim}(M;P)=\pi^* C^\infty (B)$, the cohomology class of~$\zeta$ is trivial if~$B$ is simply connected.
This setting appears in the symplectic case, where the Poisson structure~$P$ is non degenerate (see
also~\cite{MaMoRa-90}).
\end{Example}

\subsection*{Acknowledgements}

The f\/irst author (JAV) was partially supported by the Mexican Consejo Nacional de Ciencia y Tecnolog\'ia (CONACyT)
research project CB-2012 179115.
Both authors acknowledge the detailed comments of the referees, which helped to improve the contents and presentation of
this paper.

\pdfbookmark[1]{References}{ref}
\LastPageEnding


\begin{thebibliography}{99}
\footnotesize \itemsep=0pt

\bibitem{MJY-13}
Avenda{\~n}o-Camacho M., Vallejo J.A., Vorobiev Yu., Higher order corrections to
  adiabatic invariants of generalized slow-fast {H}amiltonian systems,
  \href{http://dx.doi.org/10.1063/1.4817863}{\textit{J.~Math. Phys.}} \textbf{54} (2013), 082704, 15~pages,
  \href{http://arxiv.org/abs/1305.3974}{arXiv:1305.3974}.

\bibitem{BrF}
Brahic O., Fernandes R.L., Poisson f\/ibrations and f\/ibered symplectic groupoids,
  in Poisson geometry in mathematics and physics, \href{http://dx.doi.org/10.1090/conm/450/08733}{\textit{Contemp. Math.}}, Vol.~450, Amer. Math. Soc., Providence, RI, 2008, 41--59,
  \href{http://arxiv.org/abs/math.DG/0702258}{math.DG/0702258}.

\bibitem{BrF-13}
Brahic O., Fernandes R.L., Integrability and reduction of Hamiltonian actions
  on Dirac manifolds, \href{http://arxiv.org/abs/1311.7398}{arXiv:1311.7398}.

\bibitem{Bur-05}
Bursztyn H., On gauge transformations of {P}oisson structures, in Quantum Field
  Theory and Noncommutative Geometry, \href{http://dx.doi.org/10.1007/11342786_5}{\textit{Lecture Notes in Phys.}}, Vol.~662, Springer, Berlin, 2005, 89--112.

\bibitem{Bur-Ra-03}
Bursztyn H., Radko O., Gauge equivalence of {D}irac structures and symplectic
  groupoids, \href{http://dx.doi.org/10.5802/aif.1945}{\textit{Ann. Inst. Fourier (Grenoble)}} \textbf{53} (2003),
  309--337, \href{http://arxiv.org/abs/math.SG/0202099}{math.SG/0202099}.

\bibitem{Cha-06}
Chavel I., Riemannian geometry. A~modern introduction, \href{http://dx.doi.org/10.1017/CBO9780511616822}{\textit{Cambridge
  Studies in Advanced Mathematics}}, Vol.~98, 2nd ed., Cambridge University
  Press, Cambridge, 2006.

\bibitem{Cou-90}
Courant T.J., Dirac manifolds, \href{http://dx.doi.org/10.2307/2001258}{\textit{Trans. Amer. Math. Soc.}} \textbf{319}
  (1990), 631--661.

\bibitem{CuWe-86}
Courant T.J., Weinstein A., Beyond {P}oisson structures, in Action Hamiltoniennes
  de Groupes. {T}roisi\`eme Th\'eor\`eme de {L}ie ({L}yon, 1986),
  \textit{Travaux en Cours}, Vol.~27, Hermann, Paris, 1988, 39--49.


\bibitem{CrMa-13}
Crainic M., M{\u{a}}rcu{\c{t}} I., A normal form theorem around symplectic
  leaves, \textit{J.~Differential Geom.} \textbf{92} (2012), 417--461,
  \href{http://arxiv.org/abs/1009.2090}{arXiv:1009.2090}.

\bibitem{DWa-09}
Davis B.L., Wade A., Dirac structures and gauge symmetries of phase spaces,
  \textit{Rend. Semin. Mat. Univ. Politec. Torino} \textbf{67} (2009),
  123--135.

\bibitem{DW}
Dufour J.-P., Wade A., On the local structure of {D}irac manifolds,
  \href{http://dx.doi.org/10.1112/S0010437X07003272}{\textit{Compos. Math.}} \textbf{144} (2008), 774--786,
  \href{http://arxiv.org/abs/math.SG/0405257}{math.SG/0405257}.

\bibitem{FMa-13}
Frejlich P., M\v{a}rcu\c{t} I., Poisson transversals I.~The normal form
  theorem, \href{http://arxiv.org/abs/1306.6055}{arXiv:1306.6055}.

\bibitem{JoRa-10}
Jotz M., Ratiu T.S., Induced {D}irac structures on isotropy-type manifolds,
  \href{http://dx.doi.org/10.1007/s00031-011-9123-z}{\textit{Transform. Groups}} \textbf{16} (2011), 175--191, \href{http://arxiv.org/abs/1008.2280}{arXiv:1008.2280}.

\bibitem{JRS-11}
Jotz M., Ratiu T.S., {\'S}niatycki J., Singular reduction of {D}irac
  structures, \href{http://dx.doi.org/10.1090/S0002-9947-2011-05220-7}{\textit{Trans. Amer. Math. Soc.}} \textbf{363} (2011), 2967--3013,
  \href{http://arxiv.org/abs/0901.3062}{arXiv:0901.3062}.

\bibitem{KMS-93}
Kol{\'a}{\v{r}} I., Michor P.W., Slov{\'a}k J., Natural operations in
  dif\/ferential geometry, \href{http://dx.doi.org/10.1007/978-3-662-02950-3}{Springer-Verlag}, Berlin, 1993.

\bibitem{MaMoRa-90}
Marsden J., Montgomery R., Ratiu T., Reduction, symmetry, and phases in
  mechanics, \href{http://dx.doi.org/10.1090/memo/0436}{\textit{Mem. Amer. Math. Soc.}} \textbf{88} (1990), iv+110~pages.

\bibitem{MZ-06}
Miranda E., Zung N.T., A note on equivariant normal forms of {P}oisson
  structures, \href{http://dx.doi.org/10.4310/MRL.2006.v13.n6.a14}{\textit{Math. Res. Lett.}} \textbf{13} (2006), 1001--1012,
  \href{http://arxiv.org/abs/math.SG/0510523}{math.SG/0510523}.

\bibitem{SeWe-01}
{\v{S}}evera P., Weinstein A., Poisson geometry with a 3-form background,
  \href{http://dx.doi.org/10.1143/PTPS.144.145}{\textit{Progr. Theoret. Phys. Suppl.}} \textbf{144} (2001), 145--154,
  \href{http://arxiv.org/abs/math.SG/0107133}{math.SG/0107133}.

\bibitem{Sni-13}
{\'S}niatycki J., Dif\/ferential geometry of singular spaces and reduction of
  symmetry, \href{http://dx.doi.org/10.1017/CBO9781139136990}{\textit{New Mathematical Monographs}}, Vol.~23, Cambridge University
  Press, Cambridge, 2013.

\bibitem{Ste-74}
Stefan P., Accessible sets, orbits, and foliations with singularities,
  \href{http://dx.doi.org/10.1112/plms/s3-29.4.699}{\textit{Proc. London Math. Soc.}} \textbf{29} (1974), 699--713.

\bibitem{Sus-73}
Sussmann H.J., Orbits of families of vector f\/ields and integrability of
  distributions, \href{http://dx.doi.org/10.1090/S0002-9947-1973-0321133-2}{\textit{Trans. Amer. Math. Soc.}} \textbf{180} (1973),
  171--188.

\bibitem{Va--94}
Vaisman I., Lectures on the geometry of {P}oisson manifolds, \href{http://dx.doi.org/10.1007/978-3-0348-8495-2}{\textit{Progress
  in Mathematics}}, Vol.~118, Birkh\"auser Verlag, Basel, 1994.

\bibitem{Va-04}
Vaisman I., Coupling {P}oisson and {J}acobi structures on foliated manifolds,
  \href{http://dx.doi.org/10.1142/S0219887804000307}{\textit{Int.~J. Geom. Methods Mod. Phys.}} \textbf{1} (2004), 607--637,
  \href{http://arxiv.org/abs/math.SG/0402361}{math.SG/0402361}.

\bibitem{Va-06}
Vaisman I., Foliation-coupling {D}irac structures, \href{http://dx.doi.org/10.1016/j.geomphys.2005.05.007}{\textit{J.~Geom. Phys.}}
  \textbf{56} (2006), 917--938, \href{http://arxiv.org/abs/math.SG/0412318}{math.SG/0412318}.

\bibitem{Vo-01}
Vorobjev Yu., Coupling tensors and {P}oisson geometry near a single symplectic
  leaf, in Lie Algebroids and Related Topics in Dif\/ferential Geometry
  ({W}arsaw, 2000), \textit{Banach Center Publ.}, Vol.~54, Polish Acad. Sci.
  Inst. Math., Warsaw, 2001, 249--274, \href{http://arxiv.org/abs/math.SG/0008162}{math.SG/0008162}.

\bibitem{Vo-05}
Vorobjev Yu., Poisson equivalence over a symplectic leaf, in Quantum algebras
  and {P}oisson geometry in mathematical physics, \textit{Amer. Math. Soc.
  Transl. Ser.~2}, Vol.~216, Amer. Math. Soc., Providence, RI, 2005, 241--277,
  \href{http://arxiv.org/abs/math.SG/0503628}{math.SG/0503628}.

\bibitem{Vor-08}
Vorobiev Yu., Averaging of {P}oisson structures, in Geometric Methods in
  Physics, \href{http://dx.doi.org/10.1063/1.3043864}{\textit{AIP Conf. Proc.}}, Vol.~1079, Amer. Inst. Phys., Melville,
  NY, 2008, 235--240.

\bibitem{Vor-11}
Vorobiev Yu., The averaging in Hamiltonian systems on slow-fast phase spaces
  with $\mathbb{S}^{1}$ symmetry, \href{http://dx.doi.org/10.1134/S1063778811070179}{\textit{Phys. Atomic Nuclei}} \textbf{74}
  (2011), 1770--1774.

\bibitem{Wa}
Wade A., Poisson f\/iber bundles and coupling {D}irac structures, \href{http://dx.doi.org/10.1007/s10455-007-9079-3}{\textit{Ann.
  Global Anal. Geom.}} \textbf{33} (2008), 207--217, \href{http://arxiv.org/abs/math.SG/0507594}{math.SG/0507594}.

\end{thebibliography}
\end{document}